\documentclass[iop,apj,tighten]{emulateapj}

\usepackage{graphicx,natbib,amsmath,amssymb}
\bibpunct[, ]{(}{)}{;}{a}{}{,}
\newcommand{\simgt}{\hbox{\rlap{\raise 0.425ex\hbox{$>$}}\lower 0.65ex\hbox{$\sim$}}}
\newcommand{\simlt}{\hbox{\rlap{\raise 0.425ex\hbox{$<$}}\lower 0.65ex\hbox{$\sim$}}}

\newcommand{\spitzer}{{\it Spitzer}}
\newcommand{\um}{$\mu$m}
\newcommand{\msun}{M$_\odot$}
\newcommand{\ha}{H$\alpha$}
\newcommand{\hb}{H$\beta$}
\newcommand{\oiii}{[\ion{O}{3}]}
\newcommand{\targeta}{SDSS J1030+0524}
\newcommand{\targetb}{SDSS J1044$-$0125}
\newcommand{\targetc}{SDSS J1048+4637}
\newcommand{\ratio}{$\eta$}
\newcommand{\auvx}{{\tilde A}_{1450}}
\newcommand{\auv}{{\tilde A}}
\defcitealias{2010A&A...523A..85G}{G10}

\journalinfo{The Astrophysical Journal, in press (2013)}
\submitted{Received 2013 March 4; accepted 2013 April 8}

\shorttitle{Extinction in high-redshift QSOs}
\shortauthors{Hjorth et al.}

\begin{document}

\title{
On inferring extinction laws in $z\sim6$ quasars as signatures of 
supernova dust%
}


\author{ 
Jens Hjorth\altaffilmark{1},
Paul M. Vreeswijk\altaffilmark{2,1,3},
Christa Gall\altaffilmark{4,1}, and
Darach Watson\altaffilmark{1}
}

\altaffiltext{1}{Dark Cosmology Centre, Niels Bohr Institute, University of 
Copenhagen, Juliane Maries Vej 30, DK--2100 Copenhagen \O, Denmark;
jens@dark-cosmology.dk}
\altaffiltext{2}{Department of Particle Physics and Astrophysics, Faculty of 
Physics, Weizmann Institute of Science, Rehovot 76100, Israel}
\altaffiltext{3}{Centre for Astrophysics and Cosmology, Science Institute,
University of Iceland, Dunhaga 5, IS--107 Reykjavik, Iceland}
\altaffiltext{4}{NASA, Goddard Space Flight Center, 8800 Greenbelt Road,
Greenbelt, MD 20771}


\begin{abstract}
Unusual extinction curves of high-redshift QSOs have been taken as evidence 
that dust is primarily produced by supernovae at high redshift. In particular, 
the 3000~\AA\ Todini--Ferrara--Maiolino kink in the extinction curve of the
$z=6.20$ \targetc\ has been attributed to supernova dust. Here we discuss the 
challenges in inferring robust extinction curves of high-redshift QSOs and 
critically assess previous claims of detection of supernova dust. In 
particular, we address the sensitivity to the choice of intrinsic QSO spectrum, 
the need for a long wavelength baseline, and the drawbacks in fitting 
theoretical extinction curves. In a sample of 21 QSOs at $z \sim 6$ we detect 
significant ultraviolet extinction using existing broad-band optical, 
near-infrared, and \spitzer\ photometry. The median extinction curve is 
consistent with a Small Magellanic Cloud curve with $A_{1450}\sim 0.7$ mag and 
does not exhibit any conspicuous (restframe) 2175 \AA\ or 3000 \AA\ features.
For two QSOs, \targetb\ at $z=5.78$ and \targeta\ at $z=6.31$, we further 
present \protect{X-shooter} spectra covering the wavelength range 0.9--2.5 
\um. The resulting non-parametric extinction curves do not exhibit the 
3000~\AA\ kink. Finally, in a re-analysis of literature spectra of \targetc, 
we do not find evidence for a conspicuous kink. We conclude that the existing 
evidence for a 3000~\AA\ feature is weak
and that the overall dust properties at high and low redshift show no
significant differences.
This, however,
does not preclude 
supernovae from dominating the dust budget at high redshift.
\end{abstract}



\keywords{
accretion, accretion disks ---
dust, extinction ---
galaxies: high-redshift ---
quasars: general ---
quasars: individual (\targeta, \targetb, \targetc) 
}



\section{Introduction\label{introduction}}

Characterizing the properties of dust in galaxies at high redshift is important 
for our understanding of the physics of dust formation and its interplay with 
galaxy evolution. The very large dust masses found in some high-redshift QSOs 
from sub-mm and mm detections of their far-infrared emission
\citep{2003MNRAS.344L..74P,2003A&A...406L..55B,2004MNRAS.351L..29R,
2008MNRAS.383..289P} pose a challenge to current models of the evolution of 
massive stars and galaxies. This is because the time since Big Bang at 
$z\ga 6$ ($< 1$ Gyr) is very short for large amounts ($>10^8$ \msun) of dust to 
be produced by stellar or non-stellar sources
\citep[e.g.,][]{2007ApJ...662..927D,2010A&A...522A..15M,2010ApJ...712..942M,2011A&A...528A..13G,2011A&A...528A..14G,2011A&ARv..19...43G}. 

\begin{figure}
\epsscale{1.2}
\plotone{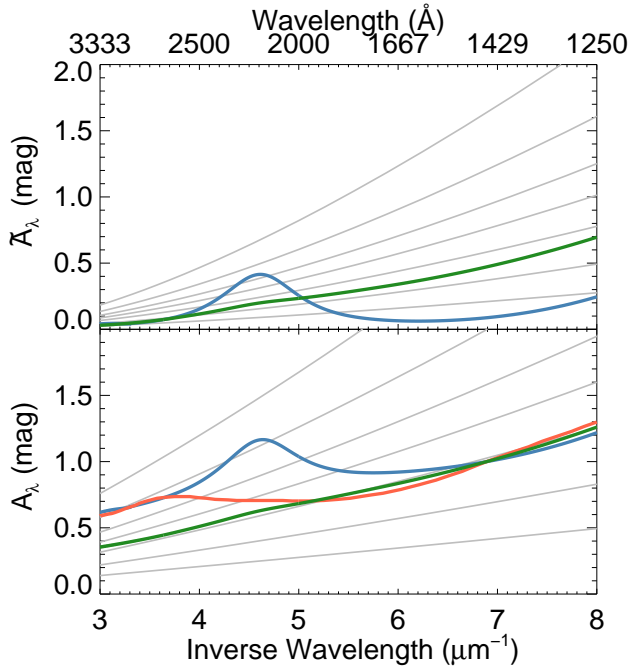}
\caption{
Absolute (bottom) and relative (top) extinction curves. Extinction curves are 
colored and normalized to $A_{1450}=1$ mag (this corresponds to $A_V=0.195$ mag 
for the SMC curve): Milky Way (blue), SMC (green), Todini--Ferrara theoretical 
extinction curve for SN dust (red). The Todini--Ferrara extinction curve is 
only plotted in the bottom panel because it is not defined above 3200 \AA. 
Artificial extinction curves are shown in gray and are not normalized. 
They are obtained from a thermal disk assumed to be a power law for 
$T^* = 40, 50, 60, 70, 85, 120, 190 \times 10^3$ K (from top to bottom), as 
discussed in Section~\ref{artificial}. Note the strong similarity to the 
SMC curve.
\label{f:disk-extinction-true}
}
\end{figure}

\citet{2004Natur.431..533M} observed the $z=6.20$ QSO \targetc\ and inferred 
an unusual ultraviolet (UV) extinction curve, inconsistent with known local 
extinction curves for the Milky Way or the Magellanic Clouds, but consistent 
with predictions of simple theoretical models of dust formation in supernova 
(SN) ejecta \citep{2001MNRAS.325..726T,2007MNRAS.378..973B} (for a variety of 
extinction curves, see Figure~\ref{f:disk-extinction-true}). The intriguing 
suggestion is that SNe are responsible for the rapid formation of dust at high 
redshift, whereas Asymptotic Giant Branch stars contribute significantly only
at lower redshift, when they have had time to evolve. 

More recently, 
\citet[][hereafter \citetalias{2010A&A...523A..85G}]{2010A&A...523A..85G} 
studied a larger sample of high-redshift QSOs, including \targetc, and
fitted their spectra with a range of theoretical SN extinction curves, 
assuming 
a QSO template spectrum \citep{2003AJ....126.2594R} with 
an adjustable power law slope 
for the underlying QSO continuum. The best-fitting 
extinction curves were found to be slightly more shallow than in the
Small Magellanic Cloud (SMC) but no strong evidence for extinction 
curves of the kind suggested by \citet{2004Natur.431..533M} was found.
However, obtaining robust results from such an approach is challenging 
because of 
(i) the narrow rest wavelength range considered ($\sim 0.13$--0.3 \um), 
(ii) the unknown intrinsic (unextinguished) QSO spectrum, 
(iii) the difficulty in estimating the continuum baseline (QSOs are 
dominated by broad and blended emission lines almost everywhere in their 
UV spectra), and 
(iv) the possible presence of variability, which will give rise to spurious 
features in the spectrum if its constituent data points were not obtained 
simultaneously.

While the prospects of interpreting unusual extinction curves as evidence 
for SN dust at high redshift are exciting, it is important to note that 
detecting significant extinction requires sufficiently blue intrinsic spectra 
of the corresponding QSOs. Indeed, previous extinction curves reported for 
high-redshift QSOs have required extremely blue QSOs. This would be consistent 
with the notion that only very blue QSOs with reddening can be detected in 
the first place because of the detection limit in the SDSS $z'$ band. However, 
it is a concern that no QSOs at high redshift with normal intrinsic colors 
have been found to have extinction, especially in view of the delicate 
procedure of inferring an extinction curve. Moreover, very blue QSOs are 
more likely to have a significant contribution from a thermal accretion disk 
\citep{2006ApJ...642...87P} which is different from the usually assumed 
intrinsic power-law spectrum. Hence, it would be important to test if 
intrinsic slopes are indeed very blue and, if so, to probe fainter QSOs 
which would allow extinction measurements in systems with normal intrinsic 
spectral slopes. 

There is only one example so far of a QSO, \targetc, with an observed 
extinction curve exhibiting a characteristic bump at around 3000 \AA\ 
\citep{2004Natur.431..533M} similar to the theoretical extinction curve of
\citet{2001MNRAS.325..726T}. \citet{2007ApJ...661L...9S} did claim evidence 
for SN dust in GRB 050904, but this was later refuted by 
\citet{2010A&A...515A..94Z}, 
as acknowledged by \citet{2011A&A...532A..45S}. \citet{2010MNRAS.406.2473P} 
and \cite{2011ApJ...741L..20J} 
have reported a \citet{2001MNRAS.325..726T} like extinction curve in 
GRB 071025, which however relies on a photometric redshift and 
hinges on the precision of the calibration of the $H$-band measurements. 
\citetalias{2010A&A...523A..85G} did interpret 
the somewhat shallower UV slopes found in their extinction curves as being 
consistent with SN dust, however, of a kind with no features in the extinction 
curves \citep{2008MNRAS.384.1725H}. It remains to be proven that an extinction 
curve with a UV slope, which is slightly shallower than the average SMC curve, 
would necessarily require SN dust, considering that radiative transfer effects 
will also give rise to shallower slopes and in view of the diversity of UV 
extinction curves in the Milky Way and the SMC 
\citep{1989ApJ...345..245C,2007ApJ...663..320F,2003ApJ...594..279G}.
Thus, observationally, unusual extinction curves are in need of confirmation. 

Regarding theoretically predicted extinction curves of SNe, there are 
considerable differences 
\citep[][\citetalias{2010A&A...523A..85G}]{2001MNRAS.325..726T,2007MNRAS.378..973B,2008MNRAS.384.1725H}. 
The underlying dust formation models are sensitive to assumptions about the 
parameters controlling the dust grain properties
\citep[such as morphology, size or composition,][]{2011A&ARv..19...43G,2011MNRAS.418..571F}    
which determine the shape and characteristics of extinction curves (see 
Section~\ref{nobumps}).  Furthermore, dust grains produced by SNe will be 
subject to either disruptive, destructive or growth processes due to, for 
example, shock interactions in the SN remnant
\citep[e.g.,][]{2007MNRAS.378..973B, 2007ApJ...666..955N, 2009MNRAS.394.1307D,2011Sci...333.1258M} 
or the interstellar medium 
\citep[e.g.,][]{2009ASPC..414..453D,2010A&A...522A..15M}, or due to 
reprocessing by the intense UV radiation in star-forming regions. Any 
modifications of the original dust grains formed by SNe will lead to changes in 
either the mineralogy or the grain-size distribution of the dust, and hence the 
resulting extinction law \citep[e.g.,][]{2011MNRAS.416.1340H}. It is therefore 
not obvious to what extent the theoretical predictions of dust extinction 
curves from SN models are directly relevant to extinction curves inferred from 
lines of sight to QSOs. Therefore, it would be prudent not to rely on specific 
parametrizations when inferring extinction curves in high-redshift QSOs. 
A non-parametric approach is desirable because we cannot assume that current 
model predictions are correct or directly applicable to observational data.

Thus motivated, this paper is devoted to a study of possible extinction signals 
in high-redshift QSOs observed by the Infrared Array Camera (IRAC) on the 
\spitzer\ {\em Space Telescope}. We first discuss the procedures of inferring 
QSO extinction, including the choice of intrinsic spectrum, specifically 
power-law models vs.\ accretion disk models (Section~2). 
Next we show that the long-wavelength IRAC/\spitzer\ data allow us 
to constrain the intrinsic spectral slopes of the QSOs and use these to 
determine rough broad-band extinction curves for the full sample of QSOs 
(Section~3). In Section~4 we present X-shooter 
observations of high-redshift QSOs, which, combined with IRAC/\spitzer\ data, 
allow us to detect extinction in high-redshift QSOs with normal colors and 
little extinction. The targets are the broad absorption line (BAL) QSO 
\targetb\ at $z=5.78$ with a large inferred dust mass and the (non-BAL) QSO 
\targeta\ at $z=6.31$ with no sub-mm detection. Dust extinction has not been 
detected in these systems previously \citepalias{2010A&A...523A..85G}. Finally, 
we study the prototypical system for a special SN extinction curve, \targetc, 
and show that there is no strong evidence for the previously claimed kink in 
the extinction curve (Section~5). We discuss our results in 
Section~6.

\section{Inferring dust extinction curves in QSOs\label{theory}}

In this section we address the problem of inferring dust extinction in QSOs 
from the wavelength-dependent attenuation of the emitted light. The standard 
method of obtaining an extinction curve is to measure the ratio between an 
extinguished spectrum and that of a reference spectrum, typically an 
unextinguished object or a model spectrum. For example, multiply imaged 
(gravitationally lensed) QSOs have been used successfully to measure 
extinction curves of intervening (lensing) galaxies 
\citep{1997A&A...317L..39J,1999ApJ...523..617F, 2000A&A...357..115T, 2002ApJ...574..719M, 2006ApJS..166..443E}
and power-law forms to measure extinction curves using 
gamma-ray burst afterglows \citep[e.g.,][]{2010A&A...515A..94Z}.

However, to measure extinction curves from the spectra of unlensed QSOs one 
must resort to comparing the observed spectrum with a model intrinsic spectrum. 
The interpretation of any signal may strongly hinge on the assumed shape of 
the intrinsic spectrum.

It is believed that a thermal accretion disk is responsible for the majority 
of the emission in the UV-optical regime of QSOs 
\citep{2005ApJ...619...41S,2006ApJ...642...87P,2008Natur.454..492K,2012MNRAS.423..451L}.
On the other hand, it is standard practise to describe the continuum emission 
in QSOs as a power law, $f_\nu \propto \nu^\alpha$, i.e.,
$f_\lambda \propto \lambda^{-2-\alpha}$, presumably resulting from reprocessing 
of the flux from the thermal disk.

A thermal disk model is defined as
\begin{equation}
f_\lambda=\left ( {6G \over c^2} \right )^2 M_{\rm BH}^2
\int_1^\infty {8 \pi^2 h c^2/\lambda^5 \over \exp (hc/(\lambda k T^* t(s))) -1} 
4 \pi s ds
\end{equation}
with $t(s) = [s^{-3}(1-s^{-1/2})]^{1/4}$ and characteristic temperature
\begin{equation}
T^*=\left ( { \dot{M}_{\rm acc} c^6 \over 576 \pi G^2 M_{\rm BH}^2 \sigma_S }\right )^{1/4}
\end{equation}
\citep{1973A&A....24..337S,1978Natur.272..706S,1992apa..book.....F,2006ApJ...642...87P}. 
In the above expressions,
$G$ is the gravitational constant,
$c$ is the speed of light,
$M_{\rm BH}$ is the black-hole mass,
$h$ is Planck's constant,
$k$ is Boltzmann's constant,
$\dot{M}_{\rm acc}$ is the mass accretion rate,
$\sigma_{\rm S}$ is the Stefan--Boltzmann constant,
and
$s$ is the radius divided by the innermost disk radius.
The latter is taken to be the last stable orbit for a non-rotating black hole, 
i.e., 3 times the Schwarzschild radius. The characteristic temperature $T^*$ 
uniquely controls the shape of the thermal disk spectrum. For $\lambda T^* \to \infty$ 
the spectrum asymptotically approaches an $\alpha=1/3$ power law. 

\begin{figure}
\epsscale{1.2}
\plotone{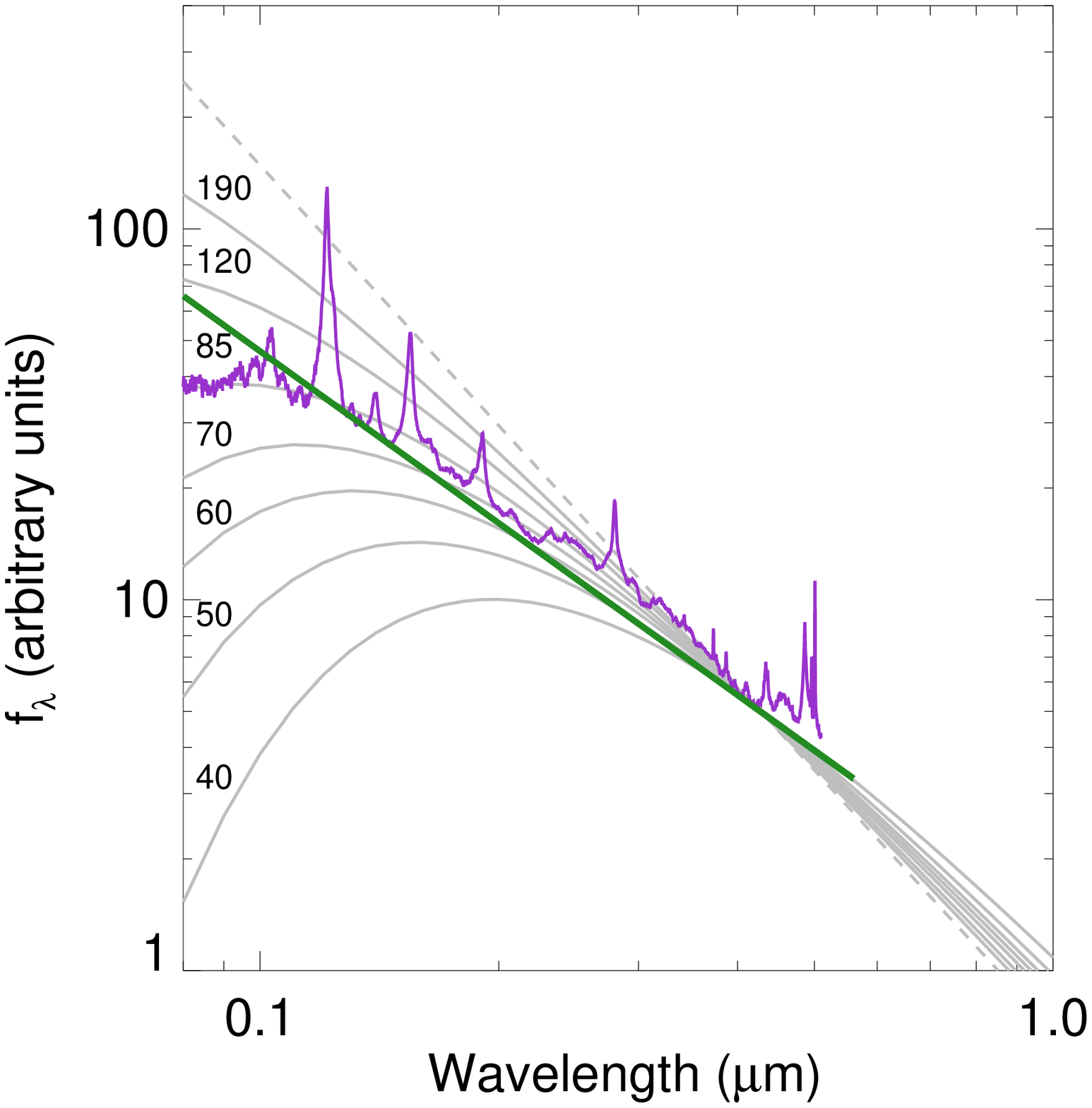}
\caption{QSO template spectrum and models. Standard thermal disks with a 
range of characteristic temperatures are plotted as gray curves
($T^* = 40, 50, 60, 70, 85, 120, 190 \times 10^3$ K from bottom to top, solid
curves and $T^* \to \infty$, dashed line). Also plotted is an {\em HST}/SDSS 
template spectrum (violet, based on \citet{2002ApJ...565..773T} and 
\citet{2001AJ....122..549V}). None of the thermal disk models provide 
acceptable representations of the template spectrum continuum. For example, 
the $T^*\approx 85000$ K model overshoots the continuum of the template 
spectrum in the range 0.13--0.23 \um. Between 0.13 \um\ and 0.50 \um\ a power 
law with spectral index $\alpha=-0.46$ (green line) is a better fit than the 
thermal disk, considering the effects of the broad emission lines 
\citep{2001AJ....122..549V}.
The SDSS template spectrum is not plotted above the location of \oiii\ 
$\lambda 5007$ where it is likely contaminated by host galaxy emission. This 
is not expected to be an issue for the high-redshift QSOs studied in this paper.
\label{f:disk}
}
\end{figure}

In Figure~\ref{f:disk} we show a thermal disk model as well as a power-law
model for the continuum of a template QSO spectrum. It is evident that a 
power-law model (with $\alpha=-0.46$) provides a better description of the 
underlying QSO spectrum (i.e., unaffected by emission lines) at wavelengths 
redward of Ly$\alpha$. A thermal disk model with $T^*\approx 85000$ K provides 
a reasonable approximation, but overshoots the continuum of the template 
spectrum in the range 0.13--0.23 \um. To avoid overshooting requires 
$T^*\la 60000$ K, but such models severely underrepresent the UV portion of 
the spectrum.

\subsection{Artificial extinction curves from accretion disk spectra modeled 
as power laws\label{artificial}}

If the assumed intrinsic model spectrum is wrong, then the method of obtaining 
extinction laws relative to this model will lead to spurious signals. For 
example, if the intrinsic spectrum is not a power law but modeled as such, 
results mimicking standard extinction laws could be inferred, even in the 
absence of dust. To quantify the effect of a thermal contribution to the 
UV-optical spectrum of a QSO we compute the artificial extinction signal 
inferred when assuming a power-law intrinsic spectral distribution for a 
continuum described by a standard thermal accretion disk. 

To obtain an effective extinction curve from a thermal model we (wrongly)
assume the thermal accretion disk to be a power law between 5150 \AA\ and 
8300 \AA\ (to mimic the analysis in the following sections) and compute the 
ratio between the accretion disk spectrum and the power law. The results are 
shown in Figure~\ref{f:disk-extinction-true}. We retrieve featureless 
`extinction laws' which are similar to, but not identical to, the SMC 
extinction curve. The inferred `extinction' is anti-correlated with the 
intrinsic temperature of the disk. This effect must be kept in mind when 
interpreting observational results although we recall that Figure~\ref{f:disk} 
demonstrates that the contribution from such a thermal spectrum cannot be 
dominant.

For the analysis of observational data, we will therefore assume a power-law 
continuum model between Ly$\alpha$ and up to 8300 \AA. This is not a good 
assumption at longer wavelengths where hot dust emission has been shown to 
dominate the emission in most high-redshift QSOs 
\citep{2006AJ....132.2127J,2010Natur.464..380J}. 
We discuss hot dust contamination in Section~\ref{spitzer-extinction}
and quantify the effect of a thermal disk contribution on our 
extinction measurements in Section~\ref{broadband-discussion}.

\section{\spitzer\ sample of $z\sim 6$ QSOs\label{spitzer-data}}

We used IRAC/\spitzer\ photometry of 21 $z\sim 6$ QSOs supplemented by
tabulated (but non-simultaneous) optical and near-infrared broad-band 
photometry published by \citet{2006AJ....132.2127J,2010Natur.464..380J}. 
The optical and near-infrared data were corrected for Galactic extinction 
\citep{1998ApJ...500..525S} assuming $R_V=3.1$. These corrections are small, 
with $E(B-V)$ ranging from 0.01 to 0.1 mag with the exception of 
SDSS J0353+0104 for which $E(B-V) = 0.29$ mag. The median value is 
$E(B-V) = 0.03$ mag. Such small Galactic extinctions are reassuring because 
the possible systematic biases on the QSO extinction signals from foreground 
extinction removal are thereby minimized.

\begin{figure}
\epsscale{1.2}
\plotone{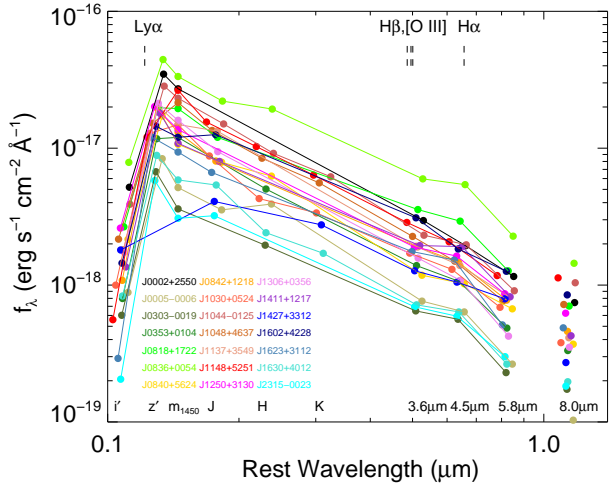}
\caption{
Optical/NIR and IRAC/\spitzer\ broad-band data for 21 $z\sim 6$ QSOs from 
\citet{2006AJ....132.2127J,2010Natur.464..380J}. The data points are not
plotted at fixed rest wavelengths because of the range in redshift 
($5.78\le z\le6.42$). Note the conspicuous effects of the emission lines of 
Ly$\alpha$ and \ha\ in the $z$ and 4.5 \um\ bands, respectively. For clarity, 
error bars are not plotted. The typical photometric uncertainties are 10--20 \% 
in the optical/NIR bands and for the 5.8 and 8.0 \um\ IRAC bands and at the 
percent level for the 3.6 and 4.5 \um\ data points.
\label{f:spitzer-SED}
}
\end{figure}

In Figure~\ref{f:spitzer-SED} we show the resulting broad-band spectral energy 
distributions. We do not plot the 24 \um\ data points because they are 
dominated by hot dust emission in most cases. The 3.6 \um, 4.5 \um, and 5.8 
\um\ data points do not seem to be affected 
\citep{2006AJ....132.2127J,2010Natur.464..380J}.
The 8.0 \um\ data points are in some cases contaminated by hot dust emission.
For this reason, these data points are not included in our analysis but shown 
in Figure~\ref{f:spitzer-SED} to illustrate the effect. Unfortunately, the 
4.5 \um\ photometry is significantly affected by \ha\ emission at $z\sim 6$ and 
the 3.6 \um\ less so by \hb\ and \oiii\ $\lambda\lambda 4959,5007$ emission 
(see Figure~\ref{f:disk}).

\subsection{Correction for emission-line 
contamination\label{spitzer-emissionline}}

\begin{figure}
\epsscale{1.2}
\plotone{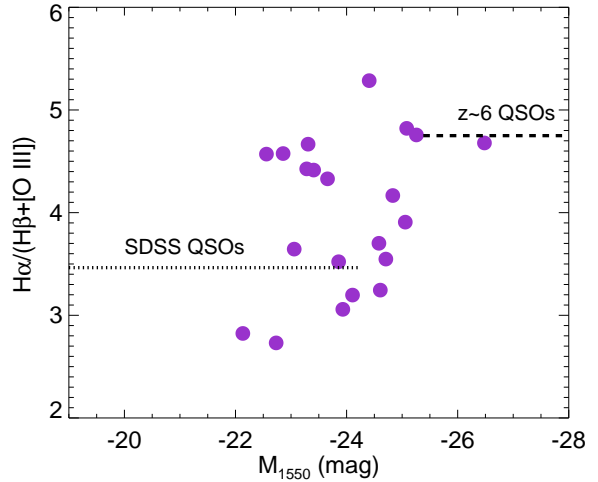}
\caption{
Relation between \ha\ to \hb+\oiii\ equivalent width ratio and luminosity
of the QSOs. The data points are for $z<0.4$ Palomar--Green QSOs
\citep{2007AJ....134..294S}. The SDSS QSO template ratio 
\citep{2001AJ....122..549V} is indicated as a dashed line spanning the range 
of SDSS QSO absolute magnitudes. The dotted line indicates the ratio adopted 
in our analysis, spanning the range of absolute magnitudes of the $z\sim6$ QSOs.
\label{f:shang}
}
\end{figure}

We correct the IRAC photometry for contamination by broad emission lines as 
follows. It is evident from Figure~\ref{f:spitzer-SED} that while there is a 
clear bump in the 4.5 \um\ band there is no conspicuous effect in the 
3.6 \um\ band. For each QSO we therefore fit a power law to the 3.6 \um\ and 
5.8 \um\ fluxes and estimate the excess \ha\ flux. The median excess flux is 
25\%, corresponding to an \ha\ equivalent width of $\sim 460$ \AA. This value 
is larger than for the template SDSS spectrum 
\citep[194 \AA;][]{2001AJ....122..549V} and consistent with the median value 
for the complete sample of $z<0.4$ Palomar--Green QSOs 
\citep[434 \AA;][]{2007AJ....134..294S}. This strongly suggests that the bump 
is indeed due to \ha\ emission and supports our method of analysis. We then 
correct the 3.6 \um\ band using an assumed equivalent width emission-line 
contamination ratio, \ratio\ $\equiv$ \ha/(\hb+\oiii). The template SDSS 
spectrum \citep{2001AJ....122..549V} suggests \ratio\ = 3.5 while the brighter 
Palomar--Green QSOs yield a median \ratio\ = 4.2, ranging from 2.7 to 5.3 
\citep{2007AJ....134..294S}, with a trend for more luminous QSOs to exhibit 
larger \ratio\ (Figure~\ref{f:shang}). These QSOs are on average an order of 
magnitude brighter than the $z<0.4$ SDSS QSOs making up the region around \ha\ 
and are therefore likely more relevant for our purposes. According to 
Figure~\ref{f:shang}, a ratio of \ratio\ = 4.75 appears to be a reasonable 
value for luminous QSOs, which we adopt in what follows. We note however that 
an even higher ratio could in principle be relevant for some of the highly 
luminous $z\sim 6$ QSOs. 

Taking into account the relative wavelength ranges of the relevant IRAC filters 
gives a ratio of 3.5 between the \ha\ to \hb+\oiii\ emission-line to continuum 
contamination, which we apply to the 3.6 \um\ flux. The procedure was repeated 
by refitting a power law to the corrected 3.6 \um\ and 5.8 \um\ fluxes until 
convergence and resulted in a median correction of $-8$\% to the 3.6 \um\ 
flux. The uncertainty in the correction induces a small systematic uncertainty 
in the inferred intrinsic power-law slope, $\alpha$, which we address below.

\subsection{Extinction\label{spitzer-extinction}}

\begin{figure}
\epsscale{1.2}
\plotone{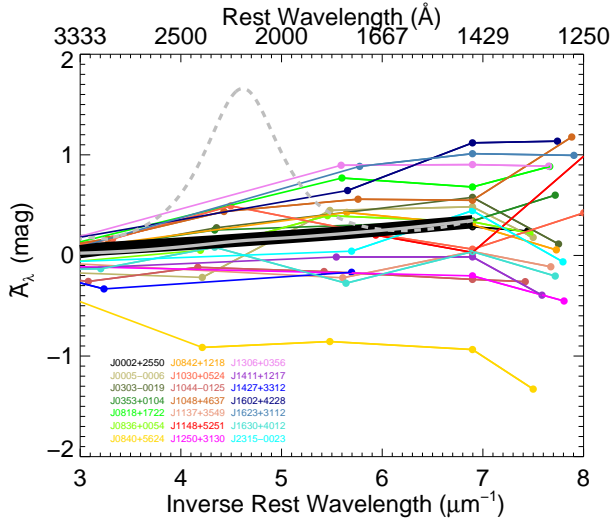}
\caption{Broad-band extinction curves relative to the IRAC/\spitzer\
power-law baseline for 21 $z\sim6$ QSOs. The thick black curve, with an 
indicative error of $0.20/\sqrt{21} = 0.04$ mag, is the median extinction 
curve of the 21 QSOs while the gray curve is a representative SMC extinction 
curve with $\auvx=0.33$ mag.  Note that for the SMC, $A_V\approx 0.41 \auvx$. 
A 2175 \AA\ bump, characteristic of the Milky Way extinction curve
(dashed gray curve),
is not seen in the median extinction curve. There is no conspicuous 
3000 \AA\ kink either.
\label{f:spitzer-ext}
}
\end{figure}

The relative extinction curve, $\auv(\lambda)$, for each QSO is obtained as the 
ratio between the restframe UV photometric data points and the extrapolated 
power-law model restframe optical spectrum, corrected for emission lines as 
explained above. Figure~\ref{f:spitzer-ext} shows the resulting extinction 
curves. There is significant scatter in the data, which are affected by other 
emission lines, depending to some extent on the actual redshift of each QSO. 
Variability may also play a role in some cases. Finally, there are substantial 
photometric uncertainties associated with the optical and near-infrared 
photometry corresponding to about 0.1--0.2 mag \citep{2010Natur.464..380J}.

Nevertheless, it is evident that the curves do not scatter around zero. If 
the data were dominated by intrinsic variability, they would scatter 
symmetrically around the zero mean. This is not the case. There is a clear 
positive effect, as expected if the QSOs are reddened relative to a power-law 
spectrum. The effect is relatively small, however, with values reaching a
maximum of about 1 mag in the restframe UV. Only one out of 21 QSOs 
(SDSS J0840+5624 at $z=5.85$) appears to exhibit a significant negative signal. 
This may be due to variability between the epochs at which the SDSS photometry, 
the near-infrared photometry, and the IRAC/\spitzer\ photometry were obtained. 
Alternatively, since we are relying on deriving a slope from just two 
literature data points, we cannot exclude problems with the reported 
photometry. In fact, the 8.0 \um\ data point for this target seems to indicate 
a somewhat steeper spectral slope which would give rise to a less negative or 
even positive extinction signal.

The data are not of sufficient quality to obtain the detailed shapes of the 
individual extinction curves. The median extinction curve is consistent with 
an SMC curve with $A_V=0.13$ mag while a Milky Way 2175 \AA\ bump is strongly 
ruled out. Likewise, there is no conspicuous 3000 \AA\ kink.

\begin{figure}
\epsscale{1.2}
\plotone{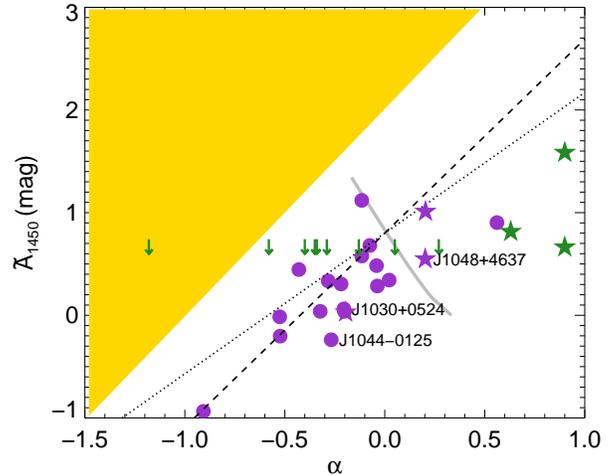}
\caption{Spectral slopes versus extinction ($\auvx$, violet symbols) for 
21 IRAC/\spitzer\ $z\sim6$ QSOs. Also shown are the results of 
\citetalias{2010A&A...523A..85G} (green symbols, stars for detections, arrows 
for upper limits). We used the \citetalias{2010A&A...523A..85G} appropriate 
mean extinction curve or the extinction curves of individual objects to
obtain $A_{1450}$ from their $A_{3000}$ and assumed the SMC extinction law 
relation $\auvx = 0.47 A_{1450}$ to obtain $\auvx$. Our corresponding results 
for the three green stars are shown as violet stars. The dashed and dotted 
lines indicate the degeneracies between $\auvx$ and $\alpha$ discussed in the 
text. The yellow region represents an exclusion zone due to the selection 
effect that reddened QSOs with a given IRAC 5.8 \um\ flux are more easily 
detected when the QSO is blue. The borderline between the excluded (yellow) 
and allowed region represents a QSO with an intrinsic $\alpha= 0$ spectrum, 
detectable at 10 $\sigma$ at 5.8 \um\ and observable to $z'=20.9$ mag in the 
SDSS. The gray line indicates the relation between excess inferred reddening 
and inferred spectral slope for a pure thermal disk, ranging from values around 
$\alpha\approx-0.2,\auvx\approx 1.3$ for low temperatures 
($T^*\approx 40000$ K) to the asymptotic values ($\alpha=0.33,\auvx=0$) at 
high temperatures.
\label{f:spitzer-A-alpha}
}
\end{figure}

For each QSO we can obtain an estimate of the UV extinction, $\auvx$, from the 
1450 \AA\ magnitude (the $z$ and $J$ bands are affected by Ly$\alpha$ and 
\ion{C}{4} emission, respectively). Figure~\ref{f:spitzer-A-alpha} shows the 
relation between the inferred extinction, $\auvx$, and spectral slope, $\alpha$.
There is a clear correlation, which can be understood as follows: For a power, 
law $f_\nu \propto \nu^\alpha$, anchored at $\lambda_{\rm anchor}$ and observed 
at $\lambda_{\rm obs}$ the difference in the inferred relative extinction at  
$\lambda_{\rm obs}$ is
\begin{equation}
\delta \auv = 2.5 \log \left ( \frac{\lambda_{\rm anchor}}{\lambda_{\rm obs}} \right )
\delta \alpha
\label{degeneracy}
\end{equation}
if the spectral slope changes by $\delta \alpha$. 

One uncertainty in $\alpha$ arises from the slightly model dependent correction 
for the flux of \hb\ and \oiii\ in the 3.6 \um\ band. The spectral slope is 
anchored at the 5.8 \um\ IRAC band, i.e., at around 8300 \AA\ in the restframe 
at $z=6$. The observed wavelength is at 1450 \AA\ in the restframe. Hence, we 
expect the inferred extinction and spectral slope data points to be degenerate 
along
\begin{equation}
\delta \auv = 1.89 \delta \alpha,
\label{degeneracy1}
\end{equation}
plotted as a dashed line in Figure~\ref{f:spitzer-A-alpha}. We find that a
variation $\delta \eta$ leads to a change in $\alpha$ of
\begin{equation}
\delta \alpha \approx 0.1 \frac{\delta \eta}{4.75},
\end{equation}
which indicates that this is a small effect.

Another source of error is the possible contamination of the 5.8 \um\ band 
by hot dust emission. If this uncertainty is dominant, the spectral slope 
is effectively anchored at the 3.6 \um\ IRAC band, i.e., at around 5150 \AA\ 
in the restframe at $z=6$, and
\begin{equation}
\delta \auv = 1.37 \delta \alpha.
\label{degeneracy2}
\end{equation}
This relation is plotted in Figure~\ref{f:spitzer-A-alpha} as a dotted line.
We find that the median value of $\alpha$ in this case varies as
\begin{equation}
\delta \alpha \approx
-1.8 \frac{\delta f_{5.8\mu{\rm m}}}{f_{5.8\mu{\rm m}}},
\end{equation}
which shows a strong sensitivity to the 5.8 \um\ flux. Fortunately, the hot 
dust contamination at 8.0 \um\ is small ($\la 10\%$, see 
Figure~\ref{f:spitzer-SED}) and hence negligible in most cases at 5.8 \um:
the Planck function drops by a factor of 9 between 8.0 \um\ and 5.8 \um\ in 
the Wien region, while the flux is roughly constant for $\alpha\approx 0$.
This is consistent with the hot dust models of \citet{2006AJ....132.2127J} 
which indicate that the contamination of the 5.8 \um\ flux is of order 1\%. 
We note that any contamination by hot dust emission will cause the 5.8 \um\ 
flux to be overestimated and hence $\alpha$ and $\auv$ to be underestimated. 
For consistency, we applied a $-1$\% correction to all 5.8 \um\ fluxes to 
account for contamination by hot dust emission.

Having shown that the systematic uncertainties related to the choice of $\eta$
or due to hot dust contamination are small, we conclude that the actual 
photometric uncertainty in the 5.8 \um\ flux dominates the error budget (the 
uncertainties in the 3.6 and 4.5 \um\ fluxes are much smaller). The median 
relative error in the 5.8 \um\ flux is 8\% which translates into a typical 
error in $\auvx$ of 0.20 mag, or 0.04 mag in the mean for the sample of 21 QSOs.

Selection effects are described by relations similar to 
Equations~(\ref{degeneracy1}) and (\ref{degeneracy2}). As the selection of 
high-redshift QSOs is based on a $z'$-filter detection and $i'$-filter dropout, 
the dominating selection effect is related to the $z'$-filter brightness. For 
a given IRAC 5.8 \um\ band detection, the corresponding constant of 
proportionality in Equation~(\ref{degeneracy1}) becomes slightly larger, 
around 2 (taking the $z'$-band to be the observed wavelength), i.e., it 
defines an exclusion zone above a line with a slope of 
$\delta \auvx /\delta \alpha = 2$ where QSOs are too faint for detection. The 
normalization depends on the intrinsic brightness of the QSOs and the depth of 
the IRAC/\spitzer\ observation. In Figure~\ref{f:spitzer-A-alpha} we illustrate 
the effect for a typical QSO detected at $100\pm10$ $\mu$Jy in the 5.8 \um\ 
band. Assuming a flat spectrum ($\alpha=0$) this corresponds to a QSO with an 
unextinguished brightness of $z'=18.9$ mag. The faintest QSO in the sample has 
$z'=20.9$ mag, i.e., we can observe 2 mag of relative extinction for a QSO with 
$\alpha=0$. \citetalias{2010A&A...523A..85G} discussed a color selection effect 
based on $z'-J$ but this constraint is weaker than the brightness constraint in 
our case and is not plotted. 

\subsection{Discussion of the broad-band results\label{broadband-discussion}}

\begin{figure}
\epsscale{1.2}
\plotone{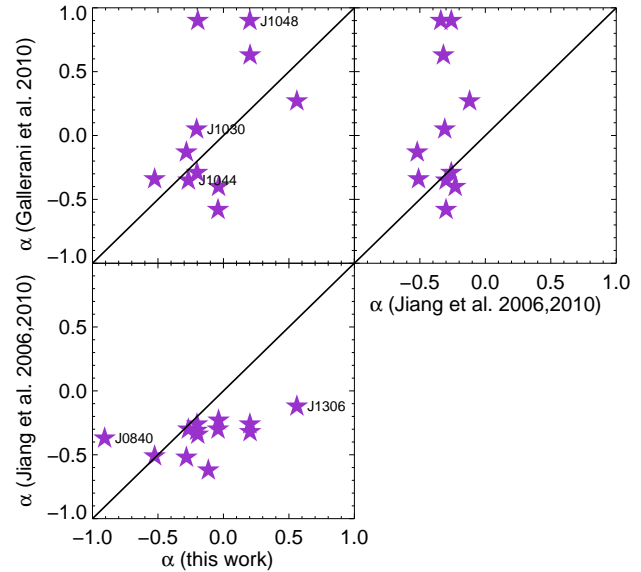}
\caption{
Comparison of inferred spectral slopes for $z\sim 6$ QSOs. We plot the 
intrinsic spectral slopes obtained in this work and compare to those of 
\citetalias{2010A&A...523A..85G} and those adopted by 
\citet{2006AJ....132.2127J,2010Natur.464..380J}. The filled stars represent 
QSOs for which $\alpha$ is reported by the different sources. The symbols 
should lie on the solid lines for consistency between the different methods 
for inferring the intrinsic spectral slopes. Note that our slopes and those 
of \citetalias{2010A&A...523A..85G} are extinction corrected whereas those of
\citet{2006AJ....132.2127J,2010Natur.464..380J} are not. This partly explains 
why the latter values are smaller, corresponding to redder sources. 
\label{f:multi-alpha}
}
\end{figure}

Figure~\ref{f:spitzer-A-alpha} illustrates the fact that available detection
limits do not allow the detection of very reddened QSOs with small spectral 
slopes (i.e., intrinsically red QSOs). Therefore, as pointed out by 
\citetalias{2010A&A...523A..85G}, the fact that previously reported 
high-redshift QSOs with significant extinction are intrinsically very blue 
appears consistent with the selection effects. On the other hand, our results 
appear to be inconsistent with the \citetalias{2010A&A...523A..85G} results. 
In Figure~\ref{f:multi-alpha} we compare the spectral slopes obtained for the 
QSOs studied here with those obtained by \citetalias{2010A&A...523A..85G}. 
There is no obvious correlation between the respective values and no systematic 
bias either. It seems the outcomes of the two approaches are very different.
Hence, given the extinction--spectral slope correlation, extinctions obtained 
in at least one of the approaches may be unreliable. In the sample of 3814 
SDSS QSOs studied by \citet{2003AJ....126.2594R}, QSOs with observed slopes of 
$\alpha\approx 1$ or bluer are extremely rare. \citet{2007ApJ...668..682D} 
find an even more restricted range of slopes in the 1450--2200 (2200--4000) 
\AA\ range from 3646 (2706) quasars with redshifts in the $z=1.67$--2.07 
(0.76--1.26) range. QSOs with $\alpha > 0.5$ are practically non-existent in 
this sample. Hence, for \citetalias{2010A&A...523A..85G} to find three 
extremely blue ($\alpha > 0.6$) $z\sim 6$ QSOs seems unlikely. None of our 
values of $\alpha$ are this extreme. Our median value of $\alpha \approx -0.2$ 
may be compared to the SDSS template slope of $-0.46$ and the mean slopes of 
$-0.59$ (1450--2200 \AA) and $-0.37$ (2200--4000 \AA) found by 
\citet{2007ApJ...668..682D}. This may indicate a slight bias towards blue 
QSOs, which may partly be due to a selection effect because even little 
extinction would make intrinsically red QSOs undetectable.

As discussed above, the main weakness of the approach adopted by 
\citetalias{2010A&A...523A..85G} is that it relies on an assumed extinction 
curve, fitted over a limited range in wavelength, to regions of the spectrum 
which may not quite reach the true continuum level. Some of these weaknesses 
are eliminated in our approach, however, at the expense of not determining 
absolute extinction values: We emphasize that the extinction curves obtained 
here are relative to the 3.6--5.8 \um\ power law model and in this sense are 
not absolute but strictly lower limits (for an SMC extension above
3000 \AA, $\auvx \approx 0.5 A_{1450}$, see Figure~\ref{f:spitzer-A-alpha}).

As such, our approach is insensitive to a possible gray (i.e., wavelength 
independent) extinction component. However, for an entirely gray extinction 
curve our approach would have led to zero relative extinction to be detected. 
The fact that there is a clear positive detection implies that the typical 
extinction curve of high-redshift QSOs is wavelength dependent, giving rise to 
reddening as expected (still under the assumption of an intrinsic power-law 
spectrum). 

To address the question of systematic errors in the derived extinction values
due to a contribution of a thermal disk we overplot the extinction--slope 
relation inferred for a non-reddened thermal accretion disk model in 
Figure~\ref{f:spitzer-A-alpha}. The curve reflects a range of characteristic 
temperatures starting at $T^*=40000$ K ($\alpha = -0.2, \auvx=1.3$ mag) and 
approaching the asymptotic value ($\alpha = 0.33, \auvx=0)$ for large $T^*$. 
Remarkably, this curve is practically perpendicular to the degeneracy 
relations, Equations~(\ref{degeneracy1}) and (\ref{degeneracy2}), so 
basically adds to the scatter in this relation. Moreover, it is evident that 
the results do not span the full range of temperatures plotted, preferring 
values in the range $T^*=60000$--100000 K. Interestingly, the fact that the 
majority of the points do not lie on this relation shows that the QSO spectra 
are not well described by pure thermal disks and require a significant 
additional component, such as the power law assumed in our analysis.

\section{X-shooter observations\label{xshooter}}

\subsection{Targets}

\subsubsection{\targeta}

\targeta\ \citep{2001AJ....122.2833F} is a $z=6.309\pm0.009$ QSO
\citep{2007AJ....134.1150J,2004ApJ...614...69I}. \citet{2003MNRAS.344L..74P} 
obtained an upper limit to the dust mass of $1.4\times 10^8$ \msun\ from SCUBA 
observations. The source is located in a crowded sub-mm field. The sub-mm 
sources in the field are, however, likely due to foreground galaxies 
\citep{2008MNRAS.383..289P}. A deep X-ray spectrum did not reveal evidence for 
X-ray absorption 
\citep[$N_H < 8\times10^{22}$ cm$^{-2}$;][]{2004ApJ...611L..13F}. 
\citet{2006AJ....132.2127J} find a hot dust mass larger than 17 \msun. There 
are Ly$\alpha$ emitters in the field at $z=5.7$, one consistent with being a 
foreground absorber \citep{2011arXiv1104.4194G} in the QSO spectrum. There is 
also a strong \ion{Mg}{2} absorber at $z=2.780$ \citep{2011ApJ...743...21S}. 
\citet{2010Natur.464..380J} report a black hole mass of $2.75\times 10^9$ \msun.

\subsubsection{\targetb}

\targetb\ \citep{2000AJ....120.1167F} is a BAL QSO 
\citep{2001A&A...372L...5M,2001ApJ...560L...5D, 2001ApJ...561L..23G, 2003ApJ...587L..67F}
at $z=5.778\pm0.004$ \citep{2007AJ....134.1150J} with a large cold dust mass 
of $4.2\times 10^8$ \msun\ \citep{2003MNRAS.344L..74P}
and a hot dust mass larger than 53 \msun\ \citep{2006AJ....132.2127J} (second 
highest in the sample). It could be magnified by at most a factor of 2 due to 
gravitational lensing \citep{2002PASJ...54..975S,2004MNRAS.351.1266W} so the 
derived properties are likely not strongly overestimated. 
\citet{2010Natur.464..380J} find a black hole mass of $10.5\times 10^9$ \msun.

\subsection{Observations\label{paul}}

We observed \targetb\ and \targeta\ on 22 and 23 December 2009 UT,
respectively, with the X-shooter spectrograph 
\citep{2006SPIE.6269E..98D,2011A&A...536A.105V} mounted at the Cassegrain 
focus of the {\it Kueyen} unit of the Very Large Telescope (VLT) at the 
European Southern Observatory (ESO), Cerro Paranal, Chile. The observations 
started at around 7 UT on both nights, and consisted of a set of $4\times1200$ 
s integrations, with the objects being offset along the slit between the 
exposures. The seeing as measured from the guide stars used for the QSO 
observations was approximately 0.7\arcsec\ for both nights. 
During the \targetb\ observations thin clouds were scattered across the sky, 
while the sky was clear during the next night when \targeta\ was observed. The 
approximate airmass range for both QSO observations was 1.1--1.3, i.e., with 
a maximum zenith angle of 40 degrees.

X-shooter allows for simultaneous spectroscopic observations at intermediate 
resolution in three different arms: the ultra-violet and blue (UVB), visual 
(VIS) and near-infrared (NIR) wavebands, with a continuous spectral coverage 
of 0.3--2.5 \um. In this paper we only report results for the VIS and NIR arms
as the UVB arm covers restframe wavelengths much smaller than Ly$\alpha$ 
(1216 \AA). We used slit widths of 0.9\arcsec\ for both the VIS and NIR arms, 
resulting in resolving powers of 8800 (VIS) and 5600 (NIR). Immediately 
following the QSO integrations, a telluric standard star was observed at a 
very similar airmass and with the same instrument set-up as the QSOs, i.e., 
with the same slit widths and binning (no binning in the spatial direction, 
and two pixels in the dispersion direction for the VIS arm). During both nights, 
observations of spectrophotometric standards were also secured. The QSOs and 
the standards were all observed with the slit aligned with the parallactic 
angle.

Using version 1.2.2 of the X-shooter pipeline 
\citep[see][]{2006SPIE.6269E..98D} in polynomial mode, the object, telluric 
and standard star spectra were reduced to two-dimensional bias-subtracted, 
flat-field corrected, order rectified and wavelength calibrated spectra in
counts. One-dimensional spectra were optimally extracted from these
\citep[see][]{1986PASP...98..609H} with a custom IDL routine.

We used the general version of the IDL-based {\it xtellcor} program developed 
by \citet{2003PASP..115..389V} for SpeX -- a 0.8--5.5 \um\ medium-resolution 
cross-dispersed spectrograph at the NASA Infrared Telescope Facility -- to 
perform a telluric correction and flux calibration of the QSO spectra.
{\it Xtellcor} uses a high-resolution model spectrum of Vega that is scaled, 
reddened and convolved with the instrumental resolution to match the observed 
spectrum of a telluric standard as close as possible, including the hydrogen 
absorption lines. Comparison between the model and observed telluric spectra 
then allows for correction of the telluric features in the object spectra, and 
also provides a flux calibration if the $B$ and $V$ magnitudes of the telluric 
standards are known. The telluric spectra that we used were both observed 
immediately following the QSO: HIP 49704 for \targetb\ and HIP 055011 for 
\targeta, at airmasses of 1.17 and 1.11, respectively; this is within 0.1 
airmass of that of the QSO spectra. Both telluric standards are classified as 
stellar type B9, and are therefore very similar in type to an A0V star that is 
optimally suited for this telluric correction method. The advantage of using 
the telluric spectrum to also perform the flux calibration -- using the Vega 
model -- is that the telluric standard was taken immediately following the 
QSO, at a similar airmass, and with the same slit width, which minimizes slit 
loss differences. 

We checked the flux calibration by applying the sensitivity function (which 
converts measured counts to absolute flux) inferred by the {\it xtellcor} 
program directly to a spectro-photometric standard star observed during the 
same night, and by comparing the resulting spectrum to tabulated flux values. 
For the \targetb\ night we used observations of BD+17 4708 at airmass 2.63, 
while for the \targeta\ night we used GD71 at airmass 1.45. These standards 
were observed with a 5\arcsec\ wide slit. On both nights, we find that the
inferred flux for these standards is a factor of 1.3 higher than the tabulated 
values, while this factor is constant across the spectra to within 5\% -- the 
approximate accuracy of the relative flux calibration. The offset of 30\% 
could partly be attributed to slit losses (the telluric standards and 
QSOs were taken with a 0.9\arcsec\ slit). 

Given these absolute flux calibration offsets, we compared the resulting 
flux-calibrated spectra for the QSOs with the $JHK$ magnitudes collected in 
Table~1 of \citet{2006AJ....132.2127J}. For \targetb\ the spectrum flux values 
are consistent with the magnitudes within their errors (0.05--0.1 mag). 
However, for \targeta\ the spectrum flux is too high by 30\% as measured from 
the $H$ and $K$ bands where the $f_{\lambda}$ is flat and void of spectral 
features. 

This difference prompted us to check if the flux of either QSO is variable in 
time. We measured the magnitude of \targeta\ from X-shooter $z'$-band 
acquisition images using aperture photometry and calibrated against the SDSS. 
We find $z'=19.81\pm0.06$ on the night of 23 December 2009, while we find 
$z'=19.93\pm0.04$ for the night of 8 February 2011 (taken from the ESO 
archive), i.e., reasonably consistent with being constant. This compares to 
$z'=20.05\pm0.10$ as reported in the discovery paper of 
\citet{2001AJ....122.2833F}. For \targetb, we measure a magnitude of 
$z'=19.27\pm0.07$ from our acquisition images, while 
\citet{2000AJ....120.1167F} report $z'=19.23\pm0.07$. No other X-shooter 
acquisition images, or a suitable series of images taken with other 
ESO instruments are available to check for flux variability of \targetb.
The broad-band photometry is consistent with a marginally significant
brightening of \targeta\ of $0.24\pm0.12$ mag between 2000 and 2009, while
\targetb\ is consistent with being constant.
These findings reflect the inconclusive spectroscopic results. Given that
we do not have information of the brightness of \targeta\ at the time of
the IRAC/\spitzer\ observations we proceed by not making any corrections due
to variability but note that a 30\% brightness correction would offset 
the resulting extinction curve higher by about 0.30 mag.

\begin{figure}
\epsscale{1.2}
\plotone{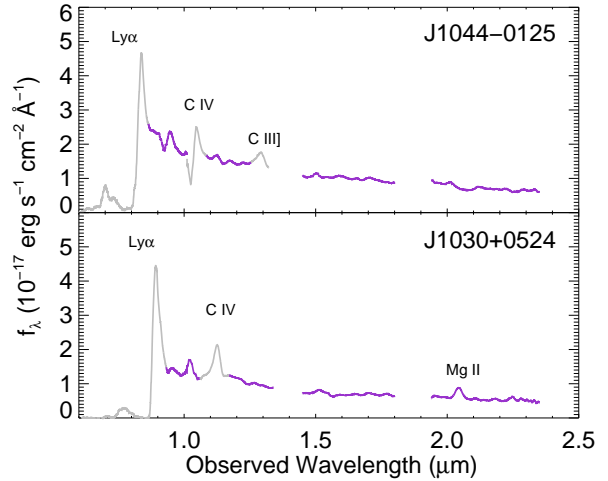}
\caption{
X-shooter/VLT spectra of \targeta\ and \targetb. The gaps in the spectra are 
due to the main atmospheric absorption bands between the $J$ and $H$ band and 
the $H$ and $K$ band. Regions plotted in gray are not included in the analysis 
due to the presence of strong emission lines (see 
Figure~\ref{f:Vandenberk-ratio}), broad absorption, or intergalactic 
absorption. The main emission lines are indicated. The spectra have been 
smoothed to 0.02 \um\ resolution.
\label{xshooter-1030-obs}
}
\end{figure}

Finally, we corrected the flux-calibrated QSO spectra for the Galactic
extinction along the sightlines, with $E(B-V)$ of 0.024 for \targeta\ and
0.051 for \targetb\, using \citet{1998ApJ...500..525S}. We display the 
X-shooter/VLT spectra in Figure~\ref{xshooter-1030-obs}.

\subsection{Extinction curves\label{xs-extinction}}

\begin{figure}
\epsscale{1.2}
\plotone{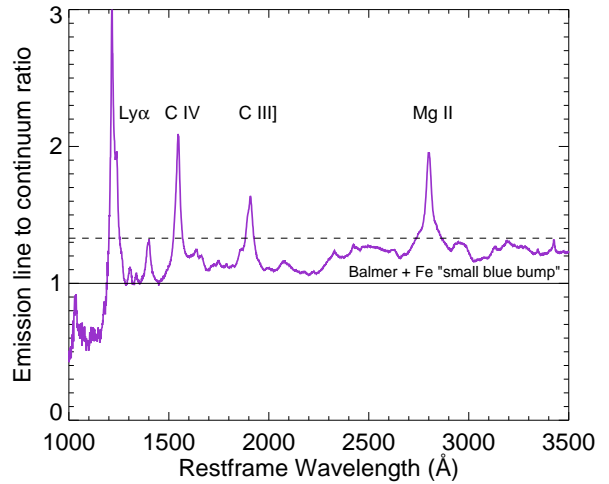}
\caption{QSO emission-line to continuum ratio obtained as the ratio between the 
QSO template and the continuum model shown in Figure~\ref{f:disk}. The solid 
line represents identity between the QSO template and the continuum. Note that 
the continuum is not reached at any point above 1450 \AA. The dashed line 
indicates regions above which the spectrum is severely affected by strong 
broad emission lines.
\label{f:Vandenberk-ratio}
}
\end{figure}

\begin{figure}
\epsscale{1.2}
\plotone{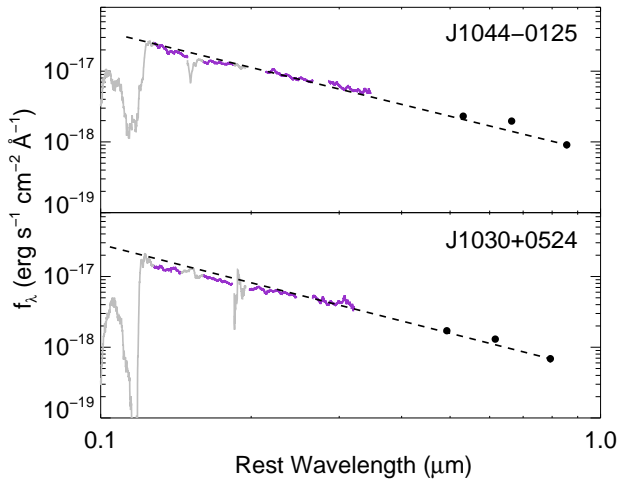}
\caption{X-shooter/VLT spectra of \targeta\ and \targetb\ after correction for
emission lines, see Figure~\ref{f:Vandenberk-ratio}. The filled circles are the 
IRAC/\spitzer\ data points. The dashed line is the adopted intrinsic power-law 
spectrum fitted to the IRAC data points as described in the text. 
The extinction is inferred as the ratio between the dashed line and the 
spectrum. As shown in Figure~\ref{xshooter-1030-ext}, the slight mismatch
between the extrapolated power laws and the observed spectra are eliminated
when choosing the 1$\sigma$ lower boundary for the 5.8 \um\ fluxes.
\label{xshooter-1030-cont}
}
\end{figure}

QSO spectra are strongly affected by emission lines, both bright conspicuous 
broad emission lines and fainter, blended lines giving rise to a 
quasi-continuum, especially around 0.2--0.3 \um, the so-called small blue 
bump due to Balmer and \ion{Fe}{2} lines. It is essential to remove the 
contamination by these lines to determine the true continuum level, assumed 
to be a power law in our analysis. We use the \citet{2001AJ....122..549V} 
SDSS QSO template to estimate the ratio between the QSO signal, affected by 
emission lines, and the continuum level (Figure~\ref{f:Vandenberk-ratio}). 
We divided the X-shooter spectra by this ratio to obtain an estimate of the 
quasi continuum, shown in Figure~\ref{xshooter-1030-cont}. It is evident that 
our procedure of removing the emission line contribution works reasonably well 
in these systems.

\begin{figure}
\epsscale{1.2}
\plotone{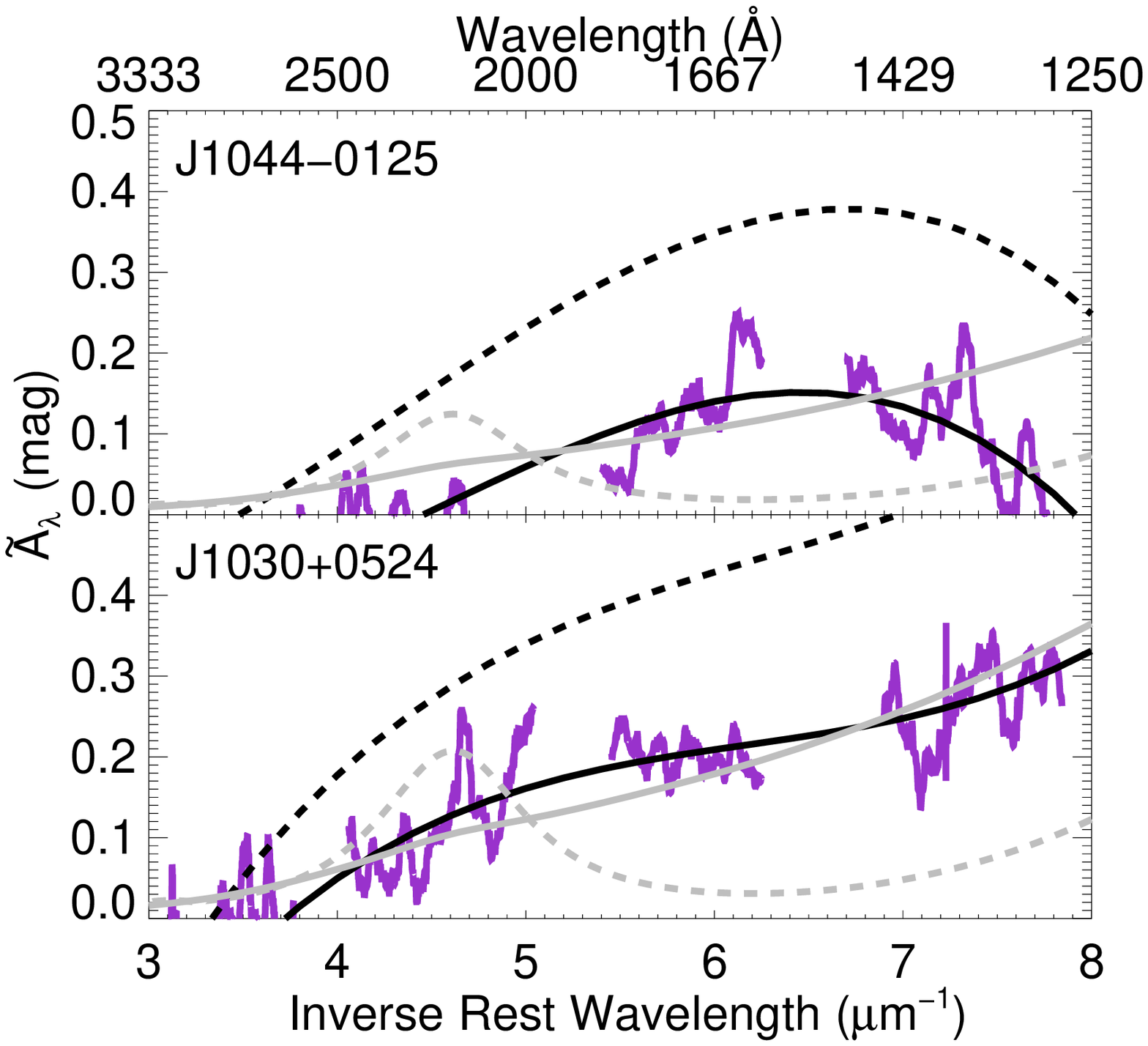}
\caption{
Inferred X-shooter/VLT extinction curves for \targeta\ and \targetb. The 
thick black line is a smooth 3rd order polynomial fit to the data. The thick 
dashed lines corresponds to assuming the lower 1$\sigma$ error on the 5.8 \um\ 
flux. If a 30\% brightness correction is applied to the \targeta\ spectrum
(see Section~\ref{paul}) the resulting extinction curve would be offset higher 
by about 0.30 mag. Indicative relative SMC (solid) and MW (dashed) extinction curves are 
overplotted in gray.
\label{xshooter-1030-ext}
}
\end{figure}

We repeated the procedure of the previous section to tie the spectra of 
\targeta\ and \targetb\ to the IRAC/\spitzer\ data. We assume that there 
has been no variability of the QSOs and that the intrinsic continuum spectrum 
is well represented by a power law as fixed by the 3.6 \um\ and 5.8 \um\ IRAC 
data points, suitably corrected for emission lines. We can then determine the 
extinction law as the ratio between the extrapolated power law and 
the emission-line corrected X-shooter spectrum. The resulting extinction 
curves are presented in Figure~\ref{xshooter-1030-ext}.

The signals are not very strong, although there seems to be the expected 
reddening effect, with $\auvx = 0.14$ mag for \targetb\ and $\auvx = 0.24$ mag 
for \targeta, consistent within the photometric errors with the broad-band 
values for these QSOs (Figures \ref{f:spitzer-ext} and \ref{f:spitzer-A-alpha}).
Neither QSO had evidence for extinction in the \citetalias{2010A&A...523A..85G} 
analysis, but their limits are fully consistent with our detections. The 
systematic errors in these extinction curves amount to $\pm \sim 0.25$ mag at 
1450 \AA\ for these systems because of the $\sim 11$\% errors in their 
5.8 \um\ fluxes. Nevertheless, we can rule out conspicuous features at 
2175 \AA\ (and to some extent 3000 \AA). 
\targeta\ has a UV rise which
is slightly shallower than the SMC, while there is an apparent
downturn in \targetb.

\section{\targetc\label{darach}}

An apparent UV-flattening was found in the extinction curve of \targetc, 
based on prism spectrophotometry using the Near Infrared Camera Spectrometer 
(NICS) instrument at the Telescopio Nazionale Galileo (TNG) and attributed to 
dust from SNe \citep{2004Natur.431..533M}. In a reanalysis of the same data,
\citetalias{2010A&A...523A..85G} found that the presence of such a 
steep-flat-steep (`kinked') extinction curve was heavily dependent on the 
intrinsic slope assumed. A different intrinsic slope resulted in
a relatively flat extinction curve without a kink. We have attempted the 
construction of extinction curves for \targetc\ using data from the 
literature, acquired using the SAO/NASA Astrophysics Data System's
Dexter data extraction applet \citep{2001ASPC..238..321D}. Optical/NIR spectra 
were acquired from \citet{2004Natur.431..533M}, \citet{2004ApJ...614...69I},
\citet{2003AJ....125.1649F,2006AJ....132..117F} based on data from NICS/TNG, 
from the OH-Airglow Suppressor/Cooled Infrared Spectrograph and Camera for OHS
(OHS/CISCO) on Subaru, from the Low-Resolution Spectrograph (LRS) on the 
Hobby-Eberly Telescope (HET), and from the Echelle Spectrograph and Imager 
(ESI) on Keck. 

\begin{figure}
\epsscale{1.2}
\plotone{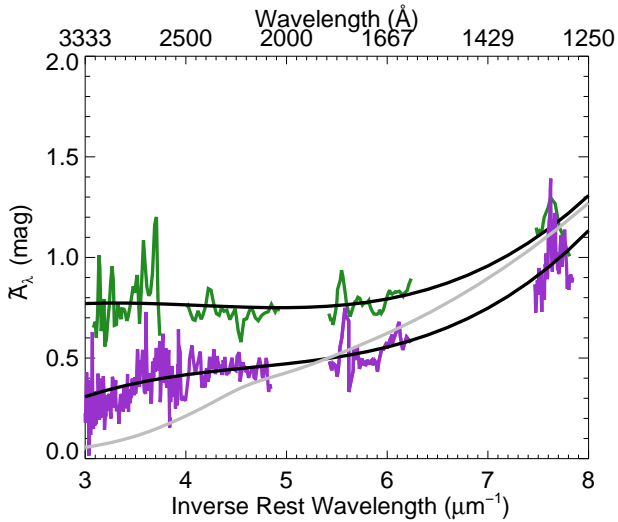}
\caption{Relative extinction curves for \targetc. The green curve is based on 
our re-analysis of the NICS/TNG data \citep{2004Natur.431..533M} whereas the 
violet curve is based on a combination of ESI/Keck and OHS/CISCO/Subaru data. 
The black overplotted curves are 3rd order polynomial fits to the data, for 
illustration. 
For comparison with the ESI+OHS curve, the relative SMC 
extinction curve is overplotted in gray. 
The curves are consistent with a flattening longward of
1700 \AA\ and a UV upturn shortward of 1700 \AA, as discussed by
\citet{2004Natur.431..533M}.
There is no conspicuous kink in 
either curves but they are clearly inconsistent with an SMC-like relative 
extinction curve.
Note that the relative G10 or \citet{2004Natur.431..533M} 
curves cannot be plotted because they are not defined above 3200 \AA. 
\label{f:J1048}
}
\end{figure}

These data were all analysed with the method described in 
Section~\ref{xs-extinction}. The resulting extinction curves using the TNG and 
the Subaru+Keck data, respectively, are shown in Figure~\ref{f:J1048} (we do
not show the LRS/HET data which are consistent with the ESI/Keck data, but of 
lower quality).

The data from NICS/TNG result in an extinction curve similar to that found by
\citet{2004Natur.431..533M}. There is however no conspicuous bump in the 
extinction curve, although a plateau is evident. In the combined 
ESI/Keck+OHS/CISCO/Subaru spectrum the overall normalization of the extinction 
curve is significantly lower and instead of a plateau there is a monotonically 
rising extinction towards the red. The shapes of the extinction curves do not 
seem very different though, with only a mild change of slope in the observed 
NIR. 
In both cases there is a UV upturn with a slope similar to the SMC. 

The differences in the derived extinction curves may be due to variability 
and/or there may be instrumental causes. In case of variability the source 
would have had to vary by as much as 60\% in the restframe UV to explain the 
difference in the absolute extinction level. Instrumental problems may play a 
role for the NICS instrument which is a very low resolution instrument, 
making sky-line subtraction complicated in the NIR. Conversely, NICS is a 
prism instrument and so is not susceptible to possible systematic slitloss 
correction errors in the way that the ESI and OHS/CISCO slit spectra may be.

We have to accept that the available data on \targetc\ can only lead us to 
inconclusive results about its extinction curve. This could be resolved with 
well-calibrated, simultaneous optical-IR spectrophotometry in the future. 
Unfortunately, the object is unobservable with X-shooter/VLT due to its 
declination.

While we confirm that the flat extinction curve is inconsistent with a typical 
SMC curve, we cannot endorse the view that there is evidence for a kinked 
`SN'-type extinction curve in this QSO, leaving no spectroscopic evidence for 
such a curve in any object to date. GRB 071025 is the only currently 
uncontested case of such an extinction curve \citep{2010MNRAS.406.2473P}. 
However, GRB 071025 has only a photometric redshift at $z\sim5$ (which impacts 
on the wavelength scale of the extinction curve) and the detection of a kink 
relies on $H$-band photometry being accurate to a few percent.

\section{Discussion\label{discussion}}

\subsection{Inferring hot dust masses}

In the process of estimating extinction in high-redshift QSOs observed by 
IRAC/\spitzer, we have devised a careful method of estimating their intrinsic 
spectral slopes. The main systematic uncertainty in this method comes from 
estimating the parameter \ratio\ which controls the correction for the emission 
lines entering the IRAC channels. Accurately estimating the intrinsic spectral 
slopes is also important for inferring hot-dust masses of these QSOs. Poor 
estimates of the spectral slope could lead to biases in the inferred
hot dust component as measured essentially in the IRAC 8.0 \um\ and 24 \um\ 
data. In Figure~\ref{f:multi-alpha} we therefore compare our spectral slopes 
to those used by \citet{2006AJ....132.2127J,2010Natur.464..380J} in their 
estimates of the mass of hot dust in the targets. It is evident that their 
spectral slopes are generally lower than ours, which leads to an underestimate 
of the hot dust signal. In particular, the hot dust mass of SDSS J1306+0356 
should probably be revised upwards because of its significantly 
higher spectral slope than that found by \citet{2006AJ....132.2127J}. 
Conversely, the hot dust mass of J0840+5624 may have to be revised downwards, 
although we note that this object was found to have spurious negative 
extinction, due to either variability or problematic photometric data 
(see Section~\ref{spitzer-extinction}).

\subsection{Correlations with other properties}

\begin{figure}
\epsscale{1.2}
\plotone{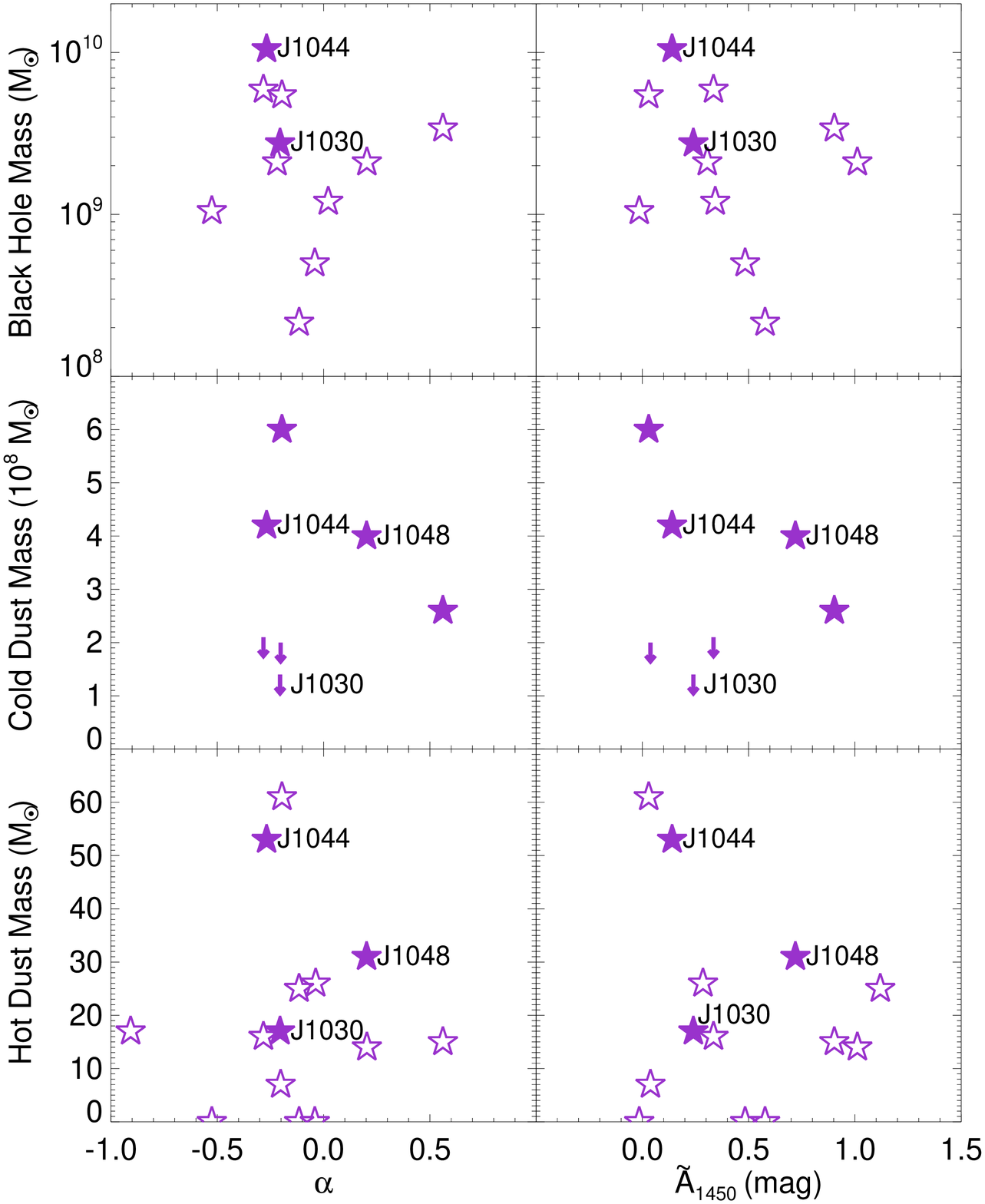}
\caption{Black hole mass, cold dust mass and hot dust mass versus spectral 
slope and extinction (violet stars). Filled symbols represent QSOs with 
sub-mm detections, arrows in the middle row represent QSOs with sub-mm upper 
limits.  The relative extinction values and spectral slopes for \targeta, 
\targetb, and \targetc\ are based on the refined analysis of their spectra 
presented in this paper. The black hole masses and hot dust masses are from 
\citet{2006AJ....132.2127J,2010Natur.464..380J} and M. Vestergaard (private 
communication). The cold dust masses are from SCUBA 850 $\mu$m observations 
\citep{2003MNRAS.344L..74P,2004MNRAS.351L..29R,2008MNRAS.383..289P}. Note that 
the two systems with $\auvx > 0.5$ mag which have been observed with SCUBA 
have detections of large masses of cold dust.
\label{m_multi}
}
\end{figure}

We compare our inferred spectral slopes and extinctions with other properties 
of the QSOs which might correlate with these values. In Figure~\ref{m_multi} 
we plot the black hole masses, cold dust masses and hot dust masses for the 
sample, to the extent that the data are available. There are no
obvious correlations between the inferred spectral slope or extinction 
with either the mass of the black hole or the mass of hot or cold dust. 
However, focusing in particular on the targets with sub-mm or mm detections or 
upper limits, Figure~\ref{m_multi} shows that the systems with large amounts of 
cold dust (the filled symbols represent systems with sub-mm detections)
also generally have large black hole masses and hot dust masses. Moreover, 
there is an indication that systems with significant extinction 
observed with SCUBA have sub-mm detections. This is encouraging and suggests 
that a significant part of the signal we find is indeed due to extinction and 
not an artifact of the data analysis or a thermal accretion disk.

\subsection{The case of \targetb}

Judging from Figure~\ref{m_multi}, \targetb\ appears to constitute a 
somewhat peculiar case because it is a BAL QSO with large amounts of inferred 
cold and hot dust and a large black hole mass. Yet, we only find evidence for 
very little extinction. We note, however, that if we assume a typical 
dust-to-gas mass ratio, and a spherical dust distribution in the host galaxy 
of the QSO, with angular diameter $<5\arcsec$ \citep{2010ApJ...714..699W}, we 
would anticipate $A_V \gtrsim 0.16$ mag for the QSO. We observe 
$\auvx\sim0.14\pm0.24$ mag which corresponds to $A_V\approx 0.06\pm0.16$ for 
an SMC-like extinction curve (see Figure~\ref{f:disk-extinction-true}). These 
values are consistent within the uncertainties. Moreover, the dust distribution 
in \targetb\ is likely not spherically distributed. For example, the galaxy 
may be flattened into a more disk-like shape. The relatively narrow width of 
the CO(6--5) line relative to the [\ion{C}{2}] linewidth 
\citep{2013arXiv1302.4154W} indicates complex gas dynamics.

\subsection{Caveats}

We seem to consistently find positive signals for the extinction and there is 
some correlation with dust signals found using other methods. Nevertheless, 
as we have stressed throughout, our approach relies on a number of assumptions, 
which we summarize below, along with additional possible systematic errors.

\begin{itemize}
\item
Likely, the most important caveat is the unknown shape of the intrinsic 
spectrum, as we have discussed at some length. The assumption of a power law 
is a convenient one, for lack of a better credible and operational model for 
the intrinsic spectrum of a QSO.
\item
We have estimated the effect of a thermal spectrum and shown that the effect 
is limited and perpendicular to the effect of extinction in the 
extinction--spectral slope plane (Figure~\ref{f:spitzer-A-alpha}). Moreover, 
we have attempted to take out the effect of emission 
lines by dividing out a template excess continuum spectrum based on SDSS QSOs
(Figure~\ref{f:Vandenberk-ratio}). On the other hand, the fake accretion disk 
extinction curves are quite reminiscent of the inferred extinction curves and 
we cannot rule out a small effect of a thermal spectrum in some of the systems. 
\item
We have determined the intrinsic power law slope using 
the IRAC/\spitzer\ detections in the 3.6 \um\ and 
5.8 \um\ bands, avoiding the 8.0 \um\ band which might be contaminated by hot 
dust emission and the 4.5 \um\ band because of \ha\ contamination. We note 
however, that the 5.8 \um\ band could in principle also be affected by hot dust 
emission in large quantities, although our analysis indicates that the effect 
will be small and only lead to an underestimate of the inferred extinction. A 
dominant contribution to the uncertainty in the extinction curves comes from 
the photometric uncertainty in the 5.8 \um\ flux.
\item
Moreover, the correction for emission line contamination of the 3.6 \um\ band 
is not trivial and we have adopted an estimate of the \ratio\ ratio from 
intermediate-redshift QSOs. A higher redshift template for more luminous 
QSOs would be highly desirable to constrain this ratio and its dispersion 
better. Fortunately, we have shown that the dependence of the inferred 
extinction on this \ratio\ ratio is weak.
\item
The IRAC/\spitzer\ data used to fix the intrinsic power law were not obtained 
simultaneously with other broad-band data or the spectroscopic data. Therefore, 
variability is a concern. Fortunately, it appears that variability is not a 
major issue for these very luminous QSOs (because of their very large black 
hole masses). Moreover, any variability is stretched in time by a factor 
$\sim 7$ due to the high redshift of the targets. However, we cannot rule out 
variability in individual cases.
\item
Simple observational aspects such as thin cirrus during the spectroscopic
observations, flux calibration difficulties, or slit loss corrections might 
also lead to overall systematic errors of order 10--30\%. Such errors could 
lead to very significant effects in the inferred extinction signal.
\item
Foreground extinction in the Galaxy is fortunately small and has been corrected 
for. However, it is always possible that there may be extinction along the 
line of sight to an object, quite unrelated to the properties of the QSO. 
Foreground extinction along random lines of sight will be about an observed 
$A_V\approx 0.1$ \citep{2010MNRAS.405.1025M}. In addition, there may be strong 
\ion{Mg}{2} absorbers along the lines of sight to individual QSOs, such as that 
detected towards \targeta\ at $z=2.78$ \citep{2011ApJ...743...21S}. In fact, 
we might expect an overdensity of \ion{Mg}{2} absorbers towards very luminous 
QSOs because of magnification due to gravitational lensing \citep[as suggested 
for \targetb, see][]{2002PASJ...54..975S}. Foreground extinction may lead to 
extinction interpreted as host galaxy extinction, and unusual extinction curves 
may arise from normal extinction curves contributing at the different 
redshifts, especially if the QSO extinction signal itself is weak.
\end{itemize}

\subsection{Theoretical SN extinction curves and the lack of evidence
for a kink\label{nobumps}}

Despite the caveats discussed above, it does appear that there are significant 
extinction signals and that these are related to the existence of dust in the 
QSO host galaxies. 

The overall shape of the obtained extinction curves primarily depend on the 
grain-size distribution, i.e., curves are typically flattened for grain-size 
distributions biased towards large grains while specific features, such as 
for example the 1700--3000 \AA\ Todini--Ferrara--Maiolino bump, depends 
on the prevalence of specific grain species. 

The `SN extinction curve' discussed in previous sections is based on the dust 
composition and grain size distribution predicted by theoretical models of 
dust formation in Type II SNe by \citet{2001MNRAS.325..726T} based on classical 
nucleation theory \citep{1966AdPhy..15..111F}. The simplifications made 
include the assumption of uniform density and temperature as well as a full 
mixing of all elements in the SN ejecta. The predominant factors shaping the 
Todini--Ferrara--Maiolino curve \citep{2004Natur.431..533M} are small 
($\approx$ 10 \AA) silicate grains (primarily forsterite Mg$_2$SiO$_4$), 
which cause the rise in the UV, and larger ($\approx$ 300 \AA) amorphous 
carbon grains. The characteristic plateau at 1700--3000 \AA\ is ascribed to 
a minimum between two broad absorption features due to amorphous carbon and a 
contribution from magnetite (Fe$_3$O$_4$) \citep{2004Natur.431..533M}. 
\citet{2007MNRAS.378..973B} modified the \citet{2001MNRAS.325..726T} dust 
formation model and included a simple analysis of dust reprocessing due to 
a reverse shock penetrating the SN ejecta. This led to a shift of the grain 
size distribution to smaller grains (dominated by magnetite), however, 
maintaining the characteristic kink mentioned above, 
as well as the UV upturn below 1700 \AA.

In the SN models for mixed and unmixed ejecta of \citet{2003ApJ...598..785N}, 
who treat several physical process in more detail, non-carbon bearing dust 
is preferentially produced, with all components having grain-size 
distributions extending to larger grains than predicted by 
\citet{2001MNRAS.325..726T}. Consequently, 
\citet{2005MNRAS.357.1077H} predict rather featureless extinction curves for 
SN dust based on these models. Extinction curves from the mixed model show a 
steep rise in the UV whereas extinction curves from the unmixed model are 
shallower. Using the same dust formation model but considering the change of 
the grain-size distribution due to the reverse shock \citep{2007ApJ...666..955N}
flattens the UV slope \citep{2008MNRAS.384.1725H} as a consequence of efficient 
destruction of small grains. Such featureless, more shallow extinction curves 
were favoured by \citetalias{2010A&A...523A..85G}, as well as by 
\citet{2011MNRAS.412.1070K} and \citet{2011MNRAS.418..625S} in $z\sim 1$ 
ultraluminous infrared galaxies. 

A common drawback of predicted SN dust extinction curves is that they are 
precariously dependent on the SN dust formation models used
\citep[for a detailed review of SN dust models, see][]{2011A&ARv..19...43G}.
The models comprise different assumptions and simplifications about complex 
physical processes of the SN evolution and dust formation leading to different 
results of dust grain properties. The size distribution is also very sensitive 
to the number of nucleation sites; more dust seeds will lead to smaller grains. 
Finally, the extinction properties are usually derived by using standard 
\citet{1908AnP...330..377M} theory for spherical grains.  

For example, \citet{2007MNRAS.378..973B} and \citet{2011MNRAS.418..571F} have 
shown that the grain-size distribution in the dust-formation models sensitively 
depends on the assumed sticking probability, which is uncertain and commonly 
adopted to be unity. The sticking probability reflects the fraction 
of the colliding condensible material which will stick together, subsequently 
building up macroscopic dust particles. Therefore, a larger sticking 
probability results in larger grains. Moreover, \citet{2011MNRAS.418..571F} 
show that aspherical carbon grains tend to be larger.

Finally, we note that none of the above predictions include the effects of the 
likely subsequent rapid reprocessing of dust in the interstellar medium of 
SN-rich galaxies
\citep[e.g.,][]{2011A&A...530A..44J,2009ASPC..414..453D,2010A&A...522A..15M,2011MNRAS.416.1340H,2012MNRAS.422.1263H,2013arXiv1303.5528A}.

In this paper we have obtained the first non-parametric extinction curves for 
dust in high-redshift QSOs. No strong evidence for unusual extinction curves, 
such as that proposed by \citet{2004Natur.431..533M}, was found. 
This suggests that the overall dust properties at low and high redshift
may be quite similar.
We stress 
however that lack of a kink or other tell-tale extinction features do 
not preclude SNe from being dominant sources of dust in these systems. Given 
the strong sensitivity to detailed SN physics and unknown dust microphysics we 
should continue to seek model independent observational evidence for SN 
extinction.

\acknowledgments
We thank Linhua Jiang and Marianne Vestergaard for providing tables of the 
broad-band photometry and the black-hole masses 
\citep[from][]{2010Natur.464..380J} used in this paper and Martin Pessah for 
helpful discussions. 
We thank 
Simona Gallerani, 
Hiroyuki Hirashita,
Roberto Maiolino,
Dan Perley,
and an anonymous referee for comments on the manuscript.
C.G. is supported from the NASA Postdoctoral Program 
(NPP). The Dark Cosmology Centre is funded by the Danish National Research 
Foundation. 

\bibliography{qsodust}

\begin{thebibliography}{81}
\expandafter\ifx\csname natexlab\endcsname\relax\def\natexlab#1{#1}\fi

\bibitem[{{Asano} {et~al.}(2013){Asano}, {Takeuchi}, {Hirashita}, \&
  {Nozawa}}]{2013arXiv1303.5528A}
{Asano}, R.~S., {Takeuchi}, T.~T., {Hirashita}, H., \& {Nozawa}, T. 2013,
  arXiv:1303.5528

\bibitem[{{Bertoldi} {et~al.}(2003){Bertoldi}, {Carilli}, {Cox}, {Fan},
  {Strauss}, {Beelen}, {Omont}, \& {Zylka}}]{2003A&A...406L..55B}
{Bertoldi}, F., {Carilli}, C.~L., {Cox}, P., {Fan}, X., {Strauss}, M.~A.,
  {Beelen}, A., {Omont}, A., \& {Zylka}, R. 2003, \aap, 406, L55

\bibitem[{{Bianchi} \& {Schneider}(2007)}]{2007MNRAS.378..973B}
{Bianchi}, S., \& {Schneider}, R. 2007, \mnras, 378, 973

\bibitem[{{Cardelli} {et~al.}(1989){Cardelli}, {Clayton}, \&
  {Mathis}}]{1989ApJ...345..245C}
{Cardelli}, J.~A., {Clayton}, G.~C., \& {Mathis}, J.~S. 1989, \apj, 345, 245

\bibitem[{{Davis} {et~al.}(2007){Davis}, {Woo}, \&
  {Blaes}}]{2007ApJ...668..682D}
{Davis}, S.~W., {Woo}, J.-H., \& {Blaes}, O.~M. 2007, \apj, 668, 682

\bibitem[{{Demleitner} {et~al.}(2001){Demleitner}, {Accomazzi}, {Eichhorn},
  {Grant}, {Kurtz}, \& {Murray}}]{2001ASPC..238..321D}
{Demleitner}, M., {Accomazzi}, A., {Eichhorn}, G., {Grant}, C.~S., {Kurtz},
  M.~J., \& {Murray}, S.~S. 2001, in Astronomical Society of the Pacific
  Conference Series, Vol. 238, Astronomical Data Analysis Software and Systems
  X, ed. F.~R. {Harnden}, Jr., F.~A. {Primini}, \& H.~E. {Payne}, 321

\bibitem[{{Djorgovski} {et~al.}(2001){Djorgovski}, {Castro}, {Stern}, \&
  {Mahabal}}]{2001ApJ...560L...5D}
{Djorgovski}, S.~G., {Castro}, S., {Stern}, D., \& {Mahabal}, A.~A. 2001,
  \apjl, 560, L5

\bibitem[{{D'Odorico} {et~al.}(2006){D'Odorico}, {Dekker}, {Mazzoleni},
  {Vernet}, {Guinouard}, {Groot}, {Hammer}, {Rasmussen}, {Kaper}, {Navarro},
  {Pallavicini}, {Peroux}, \& {Zerbi}}]{2006SPIE.6269E..98D}
{D'Odorico}, S., {et~al.} 2006, in Society of Photo-Optical Instrumentation
  Engineers (SPIE) Conference Series, Vol. 6269, Society of Photo-Optical
  Instrumentation Engineers (SPIE) Conference Series

\bibitem[{{Draine}(2009)}]{2009ASPC..414..453D}
{Draine}, B.~T. 2009, in Astronomical Society of the Pacific Conference Series,
  Vol. 414, Cosmic Dust - Near and Far, ed. T.~{Henning}, E.~{Gr{\"u}n}, \&
  J.~{Steinacker}, 453

\bibitem[{{Dunne} {et~al.}(2009){Dunne}, {Maddox}, {Ivison}, {Rudnick},
  {Delaney}, {Matthews}, {Crowe}, {Gomez}, {Eales}, \&
  {Dye}}]{2009MNRAS.394.1307D}
{Dunne}, L., {et~al.} 2009, \mnras, 394, 1307

\bibitem[{{Dwek} {et~al.}(2007){Dwek}, {Galliano}, \&
  {Jones}}]{2007ApJ...662..927D}
{Dwek}, E., {Galliano}, F., \& {Jones}, A.~P. 2007, \apj, 662, 927

\bibitem[{{El{\'{\i}}asd{\'o}ttir} {et~al.}(2006){El{\'{\i}}asd{\'o}ttir},
  {Hjorth}, {Toft}, {Burud}, \& {Paraficz}}]{2006ApJS..166..443E}
{El{\'{\i}}asd{\'o}ttir}, {\'A}., {Hjorth}, J., {Toft}, S., {Burud}, I., \&
  {Paraficz}, D. 2006, \apjs, 166, 443

\bibitem[{{Falco} {et~al.}(1999){Falco}, {Impey}, {Kochanek}, {Leh{\'a}r},
  {McLeod}, {Rix}, {Keeton}, {Mu{\~n}oz}, \& {Peng}}]{1999ApJ...523..617F}
{Falco}, E.~E., {et~al.} 1999, \apj, 523, 617

\bibitem[{{Fallest} {et~al.}(2011){Fallest}, {Nozawa}, {Nomoto}, {Umeda},
  {Maeda}, {Kozasa}, \& {Lazzati}}]{2011MNRAS.418..571F}
{Fallest}, D.~W., {Nozawa}, T., {Nomoto}, K., {Umeda}, H., {Maeda}, K.,
  {Kozasa}, T., \& {Lazzati}, D. 2011, \mnras, 418, 571

\bibitem[{{Fan} {et~al.}(2000){Fan}, {White}, {Davis}, {Becker}, {Strauss},
  {Haiman}, {Schneider}, {Gregg}, {Gunn}, {Knapp}, {Lupton}, {Anderson},
  {Anderson}, {Annis}, {Bahcall}, {Boroski}, {Brunner}, {Chen}, {Connolly},
  {Csabai}, {Doi}, {Fukugita}, {Hennessy}, {Hindsley}, {Ichikawa},
  {Ivezi{\'c}}, {Loveday}, {Meiksin}, {McKay}, {Munn}, {Newberg}, {Nichol},
  {Okamura}, {Pier}, {Sekiguchi}, {Shimasaku}, {Stoughton}, {Szalay},
  {Szokoly}, {Thakar}, {Vogeley}, \& {York}}]{2000AJ....120.1167F}
{Fan}, X., {et~al.} 2000, \aj, 120, 1167

\bibitem[{{Fan} {et~al.}(2001){Fan}, {Narayanan}, {Lupton}, {Strauss}, {Knapp},
  {Becker}, {White}, {Pentericci}, {Leggett}, {Haiman}, {Gunn}, {Ivezi{\'c}},
  {Schneider}, {Anderson}, {Brinkmann}, {Bahcall}, {Connolly}, {Csabai}, {Doi},
  {Fukugita}, {Geballe}, {Grebel}, {Harbeck}, {Hennessy}, {Lamb}, {Miknaitis},
  {Munn}, {Nichol}, {Okamura}, {Pier}, {Prada}, {Richards}, {Szalay}, \&
  {York}}]{2001AJ....122.2833F}
---. 2001, \aj, 122, 2833

\bibitem[{{Fan} {et~al.}(2003){Fan}, {Strauss}, {Schneider}, {Becker}, {White},
  {Haiman}, {Gregg}, {Pentericci}, {Grebel}, {Narayanan}, {Loh}, {Richards},
  {Gunn}, {Lupton}, {Knapp}, {Ivezi{\'c}}, {Brandt}, {Collinge}, {Hao},
  {Harbeck}, {Prada}, {Schaye}, {Strateva}, {Zakamska}, {Anderson},
  {Brinkmann}, {Bahcall}, {Lamb}, {Okamura}, {Szalay}, \&
  {York}}]{2003AJ....125.1649F}
---. 2003, \aj, 125, 1649

\bibitem[{{Fan} {et~al.}(2006){Fan}, {Strauss}, {Becker}, {White}, {Gunn},
  {Knapp}, {Richards}, {Schneider}, {Brinkmann}, \&
  {Fukugita}}]{2006AJ....132..117F}
---. 2006, \aj, 132, 117

\bibitem[{{Farrah} {et~al.}(2004){Farrah}, {Priddey}, {Wilman}, {Haehnelt}, \&
  {McMahon}}]{2004ApJ...611L..13F}
{Farrah}, D., {Priddey}, R., {Wilman}, R., {Haehnelt}, M., \& {McMahon}, R.
  2004, \apjl, 611, L13

\bibitem[{{Feder} {et~al.}(1966){Feder}, {Russell}, {Lothe}, \&
  {Pound}}]{1966AdPhy..15..111F}
{Feder}, J., {Russell}, K.~C., {Lothe}, J., \& {Pound}, G.~M. 1966, Advances in
  Physics, 15, 111

\bibitem[{{Fitzpatrick} \& {Massa}(2007)}]{2007ApJ...663..320F}
{Fitzpatrick}, E.~L., \& {Massa}, D. 2007, \apj, 663, 320

\bibitem[{{Frank} {et~al.}(1992){Frank}, {King}, \&
  {Raine}}]{1992apa..book.....F}
{Frank}, J., {King}, A., \& {Raine}, D. 1992, {Accretion power in
  astrophysics.}

\bibitem[{{Freudling} {et~al.}(2003){Freudling}, {Corbin}, \&
  {Korista}}]{2003ApJ...587L..67F}
{Freudling}, W., {Corbin}, M.~R., \& {Korista}, K.~T. 2003, \apjl, 587, L67

\bibitem[{{Gall} {et~al.}(2011{\natexlab{a}}){Gall}, {Andersen}, \&
  {Hjorth}}]{2011A&A...528A..13G}
{Gall}, C., {Andersen}, A.~C., \& {Hjorth}, J. 2011{\natexlab{a}}, \aap, 528,
  A13

\bibitem[{{Gall} {et~al.}(2011{\natexlab{b}}){Gall}, {Andersen}, \&
  {Hjorth}}]{2011A&A...528A..14G}
---. 2011{\natexlab{b}}, \aap, 528, A14

\bibitem[{{Gall} {et~al.}(2011{\natexlab{c}}){Gall}, {Hjorth}, \&
  {Andersen}}]{2011A&ARv..19...43G}
{Gall}, C., {Hjorth}, J., \& {Andersen}, A.~C. 2011{\natexlab{c}}, \aapr, 19,
  43

\bibitem[{{Gallerani} {et~al.}(2010){Gallerani}, {Maiolino}, {Juarez}, {Nagao},
  {Marconi}, {Bianchi}, {Schneider}, {Mannucci}, {Oliva}, {Willott}, {Jiang},
  \& {Fan}}]{2010A&A...523A..85G}
{Gallerani}, S., {et~al.} 2010, \aap, 523, A85

\bibitem[{{Gonzalo Diaz} {et~al.}(2011){Gonzalo Diaz}, {Ryan-Weber}, {Cooke},
  {Pettini}, \& {Madau}}]{2011arXiv1104.4194G}
{Gonzalo Diaz}, C., {Ryan-Weber}, E.~V., {Cooke}, J., {Pettini}, M., \&
  {Madau}, P. 2011, arXiv:1104.4194

\bibitem[{{Goodrich} {et~al.}(2001){Goodrich}, {Campbell}, {Chaffee}, {Hill},
  {Sprayberry}, {Brandt}, {Schneider}, {Kaspi}, {Fan}, {Gunn}, \&
  {Strauss}}]{2001ApJ...561L..23G}
{Goodrich}, R.~W., {et~al.} 2001, \apjl, 561, L23

\bibitem[{{Gordon} {et~al.}(2003){Gordon}, {Clayton}, {Misselt}, {Landolt}, \&
  {Wolff}}]{2003ApJ...594..279G}
{Gordon}, K.~D., {Clayton}, G.~C., {Misselt}, K.~A., {Landolt}, A.~U., \&
  {Wolff}, M.~J. 2003, \apj, 594, 279

\bibitem[{{Hirashita}(2012)}]{2012MNRAS.422.1263H}
{Hirashita}, H. 2012, \mnras, 422, 1263

\bibitem[{{Hirashita} \& {Kuo}(2011)}]{2011MNRAS.416.1340H}
{Hirashita}, H., \& {Kuo}, T.-M. 2011, \mnras, 416, 1340

\bibitem[{{Hirashita} {et~al.}(2005){Hirashita}, {Nozawa}, {Kozasa}, {Ishii},
  \& {Takeuchi}}]{2005MNRAS.357.1077H}
{Hirashita}, H., {Nozawa}, T., {Kozasa}, T., {Ishii}, T.~T., \& {Takeuchi},
  T.~T. 2005, \mnras, 357, 1077

\bibitem[{{Hirashita} {et~al.}(2008){Hirashita}, {Nozawa}, {Takeuchi}, \&
  {Kozasa}}]{2008MNRAS.384.1725H}
{Hirashita}, H., {Nozawa}, T., {Takeuchi}, T.~T., \& {Kozasa}, T. 2008, \mnras,
  384, 1725

\bibitem[{{Horne}(1986)}]{1986PASP...98..609H}
{Horne}, K. 1986, \pasp, 98, 609

\bibitem[{{Iwamuro} {et~al.}(2004){Iwamuro}, {Kimura}, {Eto}, {Maihara},
  {Motohara}, {Yoshii}, \& {Doi}}]{2004ApJ...614...69I}
{Iwamuro}, F., {Kimura}, M., {Eto}, S., {Maihara}, T., {Motohara}, K.,
  {Yoshii}, Y., \& {Doi}, M. 2004, \apj, 614, 69

\bibitem[{{Jang} {et~al.}(2011){Jang}, {Im}, {Lee}, {Urata}, {Huang},
  {Hirashita}, {Fan}, \& {Jiang}}]{2011ApJ...741L..20J}
{Jang}, M., {Im}, M., {Lee}, I., {Urata}, Y., {Huang}, K., {Hirashita}, H.,
  {Fan}, X., \& {Jiang}, L. 2011, \apjl, 741, L20

\bibitem[{{Jaunsen} \& {Hjorth}(1997)}]{1997A&A...317L..39J}
{Jaunsen}, A.~O., \& {Hjorth}, J. 1997, \aap, 317, L39

\bibitem[{{Jiang} {et~al.}(2007){Jiang}, {Fan}, {Vestergaard}, {Kurk},
  {Walter}, {Kelly}, \& {Strauss}}]{2007AJ....134.1150J}
{Jiang}, L., {Fan}, X., {Vestergaard}, M., {Kurk}, J.~D., {Walter}, F.,
  {Kelly}, B.~C., \& {Strauss}, M.~A. 2007, \aj, 134, 1150

\bibitem[{{Jiang} {et~al.}(2006){Jiang}, {Fan}, {Hines}, {Shi}, {Vestergaard},
  {Bertoldi}, {Brandt}, {Carilli}, {Cox}, {Le Floc'h}, {Pentericci},
  {Richards}, {Rieke}, {Schneider}, {Strauss}, {Walter}, \&
  {Brinkmann}}]{2006AJ....132.2127J}
{Jiang}, L., {et~al.} 2006, \aj, 132, 2127

\bibitem[{{Jiang} {et~al.}(2010){Jiang}, {Fan}, {Brandt}, {Carilli}, {Egami},
  {Hines}, {Kurk}, {Richards}, {Shen}, {Strauss}, {Vestergaard}, \&
  {Walter}}]{2010Natur.464..380J}
---. 2010, \nat, 464, 380

\bibitem[{{Jones} \& {Nuth}(2011)}]{2011A&A...530A..44J}
{Jones}, A.~P., \& {Nuth}, J.~A. 2011, \aap, 530, A44

\bibitem[{{Kawara} {et~al.}(2011){Kawara}, {Hirashita}, {Nozawa}, {Kozasa},
  {Oyabu}, {Matsuoka}, {Shimizu}, {Sameshima}, \&
  {Ienaka}}]{2011MNRAS.412.1070K}
{Kawara}, K., {et~al.} 2011, \mnras, 412, 1070

\bibitem[{{Kishimoto} {et~al.}(2008){Kishimoto}, {Antonucci}, {Blaes},
  {Lawrence}, {Boisson}, {Albrecht}, \& {Leipski}}]{2008Natur.454..492K}
{Kishimoto}, M., {Antonucci}, R., {Blaes}, O., {Lawrence}, A., {Boisson}, C.,
  {Albrecht}, M., \& {Leipski}, C. 2008, \nat, 454, 492

\bibitem[{{Lawrence}(2012)}]{2012MNRAS.423..451L}
{Lawrence}, A. 2012, \mnras, 423, 451

\bibitem[{{Maiolino} {et~al.}(2001){Maiolino}, {Mannucci}, {Baffa}, {Gennari},
  \& {Oliva}}]{2001A&A...372L...5M}
{Maiolino}, R., {Mannucci}, F., {Baffa}, C., {Gennari}, S., \& {Oliva}, E.
  2001, \aap, 372, L5

\bibitem[{{Maiolino} {et~al.}(2004){Maiolino}, {Schneider}, {Oliva}, {Bianchi},
  {Ferrara}, {Mannucci}, {Pedani}, \& {Roca Sogorb}}]{2004Natur.431..533M}
{Maiolino}, R., {Schneider}, R., {Oliva}, E., {Bianchi}, S., {Ferrara}, A.,
  {Mannucci}, F., {Pedani}, M., \& {Roca Sogorb}, M. 2004, \nat, 431, 533

\bibitem[{{Matsuura} {et~al.}(2011){Matsuura}, {Dwek}, {Meixner}, {Otsuka},
  {Babler}, {Barlow}, {Roman-Duval}, {Engelbracht}, {Sandstrom},
  {Laki{\'c}evi{\'c}}, {van Loon}, {Sonneborn}, {Clayton}, {Long}, {Lundqvist},
  {Nozawa}, {Gordon}, {Hony}, {Panuzzo}, {Okumura}, {Misselt}, {Montiel}, \&
  {Sauvage}}]{2011Sci...333.1258M}
{Matsuura}, M., {et~al.} 2011, Science, 333, 1258

\bibitem[{{M{\'e}nard} {et~al.}(2010){M{\'e}nard}, {Scranton}, {Fukugita}, \&
  {Richards}}]{2010MNRAS.405.1025M}
{M{\'e}nard}, B., {Scranton}, R., {Fukugita}, M., \& {Richards}, G. 2010,
  \mnras, 405, 1025

\bibitem[{{Micha{\l}owski} {et~al.}(2010{\natexlab{a}}){Micha{\l}owski},
  {Murphy}, {Hjorth}, {Watson}, {Gall}, \& {Dunlop}}]{2010A&A...522A..15M}
{Micha{\l}owski}, M.~J., {Murphy}, E.~J., {Hjorth}, J., {Watson}, D., {Gall},
  C., \& {Dunlop}, J.~S. 2010{\natexlab{a}}, \aap, 522, A15

\bibitem[{{Micha{\l}owski} {et~al.}(2010{\natexlab{b}}){Micha{\l}owski},
  {Watson}, \& {Hjorth}}]{2010ApJ...712..942M}
{Micha{\l}owski}, M.~J., {Watson}, D., \& {Hjorth}, J. 2010{\natexlab{b}},
  \apj, 712, 942

\bibitem[{{Mie}(1908)}]{1908AnP...330..377M}
{Mie}, G. 1908, Annalen der Physik, 330, 377

\bibitem[{{Motta} {et~al.}(2002){Motta}, {Mediavilla}, {Mu{\~n}oz}, {Falco},
  {Kochanek}, {Arribas}, {Garc{\'{\i}}a-Lorenzo}, {Oscoz}, \&
  {Serra-Ricart}}]{2002ApJ...574..719M}
{Motta}, V., {et~al.} 2002, \apj, 574, 719

\bibitem[{{Nozawa} {et~al.}(2007){Nozawa}, {Kozasa}, {Habe}, {Dwek}, {Umeda},
  {Tominaga}, {Maeda}, \& {Nomoto}}]{2007ApJ...666..955N}
{Nozawa}, T., {Kozasa}, T., {Habe}, A., {Dwek}, E., {Umeda}, H., {Tominaga},
  N., {Maeda}, K., \& {Nomoto}, K. 2007, \apj, 666, 955

\bibitem[{{Nozawa} {et~al.}(2003){Nozawa}, {Kozasa}, {Umeda}, {Maeda}, \&
  {Nomoto}}]{2003ApJ...598..785N}
{Nozawa}, T., {Kozasa}, T., {Umeda}, H., {Maeda}, K., \& {Nomoto}, K. 2003,
  \apj, 598, 785

\bibitem[{{Pereyra} {et~al.}(2006){Pereyra}, {Vanden Berk}, {Turnshek},
  {Hillier}, {Wilhite}, {Kron}, {Schneider}, \&
  {Brinkmann}}]{2006ApJ...642...87P}
{Pereyra}, N.~A., {Vanden Berk}, D.~E., {Turnshek}, D.~A., {Hillier}, D.~J.,
  {Wilhite}, B.~C., {Kron}, R.~G., {Schneider}, D.~P., \& {Brinkmann}, J. 2006,
  \apj, 642, 87

\bibitem[{{Perley} {et~al.}(2010){Perley}, {Bloom}, {Klein}, {Covino},
  {Minezaki}, {Wo{\'z}niak}, {Vestrand}, {Williams}, {Milne}, {Butler},
  {Updike}, {Kr{\"u}hler}, {Afonso}, {Antonelli}, {Cowie}, {Ferrero},
  {Greiner}, {Hartmann}, {Kakazu}, {K{\"u}pc{\"u} Yolda{\c s}}, {Morgan},
  {Price}, {Prochaska}, \& {Yoshii}}]{2010MNRAS.406.2473P}
{Perley}, D.~A., {et~al.} 2010, \mnras, 406, 2473

\bibitem[{{Priddey} {et~al.}(2003){Priddey}, {Isaak}, {McMahon}, {Robson}, \&
  {Pearson}}]{2003MNRAS.344L..74P}
{Priddey}, R.~S., {Isaak}, K.~G., {McMahon}, R.~G., {Robson}, E.~I., \&
  {Pearson}, C.~P. 2003, \mnras, 344, L74

\bibitem[{{Priddey} {et~al.}(2008){Priddey}, {Ivison}, \&
  {Isaak}}]{2008MNRAS.383..289P}
{Priddey}, R.~S., {Ivison}, R.~J., \& {Isaak}, K.~G. 2008, \mnras, 383, 289

\bibitem[{{Reichard} {et~al.}(2003){Reichard}, {Richards}, {Hall}, {Schneider},
  {Vanden Berk}, {Fan}, {York}, {Knapp}, \& {Brinkmann}}]{2003AJ....126.2594R}
{Reichard}, T.~A., {et~al.} 2003, \aj, 126, 2594

\bibitem[{{Robson} {et~al.}(2004){Robson}, {Priddey}, {Isaak}, \&
  {McMahon}}]{2004MNRAS.351L..29R}
{Robson}, I., {Priddey}, R.~S., {Isaak}, K.~G., \& {McMahon}, R.~G. 2004,
  \mnras, 351, L29

\bibitem[{{Schlegel} {et~al.}(1998){Schlegel}, {Finkbeiner}, \&
  {Davis}}]{1998ApJ...500..525S}
{Schlegel}, D.~J., {Finkbeiner}, D.~P., \& {Davis}, M. 1998, \apj, 500, 525

\bibitem[{{Shakura} \& {Sunyaev}(1973)}]{1973A&A....24..337S}
{Shakura}, N.~I., \& {Sunyaev}, R.~A. 1973, \aap, 24, 337

\bibitem[{{Shang} {et~al.}(2007){Shang}, {Wills}, {Wills}, \&
  {Brotherton}}]{2007AJ....134..294S}
{Shang}, Z., {Wills}, B.~J., {Wills}, D., \& {Brotherton}, M.~S. 2007, \aj,
  134, 294

\bibitem[{{Shang} {et~al.}(2005){Shang}, {Brotherton}, {Green}, {Kriss},
  {Scott}, {Quijano}, {Blaes}, {Hubeny}, {Hutchings}, {Kaiser}, {Koratkar},
  {Oegerle}, \& {Zheng}}]{2005ApJ...619...41S}
{Shang}, Z., {et~al.} 2005, \apj, 619, 41

\bibitem[{{Shields}(1978)}]{1978Natur.272..706S}
{Shields}, G.~A. 1978, \nat, 272, 706

\bibitem[{{Shimizu} {et~al.}(2011){Shimizu}, {Kawara}, {Sameshima}, {Ienaka},
  {Nozawa}, \& {Kozasa}}]{2011MNRAS.418..625S}
{Shimizu}, T., {Kawara}, K., {Sameshima}, H., {Ienaka}, N., {Nozawa}, T., \&
  {Kozasa}, T. 2011, \mnras, 418, 625

\bibitem[{{Shioya} {et~al.}(2002){Shioya}, {Taniguchi}, {Murayama}, {Ajiki},
  {Nagao}, {Fujita}, {Kakazu}, {Komiyama}, {Okamura}, {Oyabu}, {Kawara},
  {Ohyama}, {Kawabata}, {Ando}, {Nishimura}, {Hayashi}, {Ogasawara}, \&
  {Ichikawa}}]{2002PASJ...54..975S}
{Shioya}, Y., {et~al.} 2002, \pasj, 54, 975

\bibitem[{{Simcoe} {et~al.}(2011){Simcoe}, {Cooksey}, {Matejek}, {Burgasser},
  {Bochanski}, {Lovegrove}, {Bernstein}, {Pipher}, {Forrest}, {McMurtry},
  {Fan}, \& {O'Meara}}]{2011ApJ...743...21S}
{Simcoe}, R.~A., {et~al.} 2011, \apj, 743, 21

\bibitem[{{Stratta} {et~al.}(2011){Stratta}, {Gallerani}, \&
  {Maiolino}}]{2011A&A...532A..45S}
{Stratta}, G., {Gallerani}, S., \& {Maiolino}, R. 2011, \aap, 532, A45

\bibitem[{{Stratta} {et~al.}(2007){Stratta}, {Maiolino}, {Fiore}, \&
  {D'Elia}}]{2007ApJ...661L...9S}
{Stratta}, G., {Maiolino}, R., {Fiore}, F., \& {D'Elia}, V. 2007, \apjl, 661,
  L9

\bibitem[{{Telfer} {et~al.}(2002){Telfer}, {Zheng}, {Kriss}, \&
  {Davidsen}}]{2002ApJ...565..773T}
{Telfer}, R.~C., {Zheng}, W., {Kriss}, G.~A., \& {Davidsen}, A.~F. 2002, \apj,
  565, 773

\bibitem[{{Todini} \& {Ferrara}(2001)}]{2001MNRAS.325..726T}
{Todini}, P., \& {Ferrara}, A. 2001, \mnras, 325, 726

\bibitem[{{Toft} {et~al.}(2000){Toft}, {Hjorth}, \&
  {Burud}}]{2000A&A...357..115T}
{Toft}, S., {Hjorth}, J., \& {Burud}, I. 2000, \aap, 357, 115

\bibitem[{{Vacca} {et~al.}(2003){Vacca}, {Cushing}, \&
  {Rayner}}]{2003PASP..115..389V}
{Vacca}, W.~D., {Cushing}, M.~C., \& {Rayner}, J.~T. 2003, \pasp, 115, 389

\bibitem[{{Vanden Berk} {et~al.}(2001){Vanden Berk}, {Richards}, {Bauer},
  {Strauss}, {Schneider}, {Heckman}, {York}, {Hall}, {Fan}, {Knapp},
  {Anderson}, {Annis}, {Bahcall}, {Bernardi}, {Briggs}, {Brinkmann}, {Brunner},
  {Burles}, {Carey}, {Castander}, {Connolly}, {Crocker}, {Csabai}, {Doi},
  {Finkbeiner}, {Friedman}, {Frieman}, {Fukugita}, {Gunn}, {Hennessy},
  {Ivezi{\'c}}, {Kent}, {Kunszt}, {Lamb}, {Leger}, {Long}, {Loveday}, {Lupton},
  {Meiksin}, {Merelli}, {Munn}, {Newberg}, {Newcomb}, {Nichol}, {Owen}, {Pier},
  {Pope}, {Rockosi}, {Schlegel}, {Siegmund}, {Smee}, {Snir}, {Stoughton},
  {Stubbs}, {SubbaRao}, {Szalay}, {Szokoly}, {Tremonti}, {Uomoto}, {Waddell},
  {Yanny}, \& {Zheng}}]{2001AJ....122..549V}
{Vanden Berk}, D.~E., {et~al.} 2001, \aj, 122, 549

\bibitem[{{Vernet} {et~al.}(2011){Vernet}, {Dekker}, {D'Odorico}, {Kaper},
  {Kjaergaard}, {Hammer}, {Randich}, {Zerbi}, {Groot}, {Hjorth}, {Guinouard},
  {Navarro}, {Adolfse}, {Albers}, {Amans}, {Andersen}, {Andersen}, {Binetruy},
  {Bristow}, {Castillo}, {Chemla}, {Christensen}, {Conconi}, {Conzelmann},
  {Dam}, {de Caprio}, {de Ugarte Postigo}, {Delabre}, {di Marcantonio},
  {Downing}, {Elswijk}, {Finger}, {Fischer}, {Flores}, {Fran{\c c}ois},
  {Goldoni}, {Guglielmi}, {Haigron}, {Hanenburg}, {Hendriks}, {Horrobin},
  {Horville}, {Jessen}, {Kerber}, {Kern}, {Kiekebusch}, {Kleszcz}, {Klougart},
  {Kragt}, {Larsen}, {Lizon}, {Lucuix}, {Mainieri}, {Manuputy}, {Martayan},
  {Mason}, {Mazzoleni}, {Michaelsen}, {Modigliani}, {Moehler}, {M{\o}ller},
  {Norup S{\o}rensen}, {N{\o}rregaard}, {P{\'e}roux}, {Patat}, {Pena}, {Pragt},
  {Reinero}, {Rigal}, {Riva}, {Roelfsema}, {Royer}, {Sacco}, {Santin},
  {Schoenmaker}, {Spano}, {Sweers}, {Ter Horst}, {Tintori}, {Tromp}, {van
  Dael}, {van der Vliet}, {Venema}, {Vidali}, {Vinther}, {Vola}, {Winters},
  {Wistisen}, {Wulterkens}, \& {Zacchei}}]{2011A&A...536A.105V}
{Vernet}, J., {et~al.} 2011, \aap, 536, A105

\bibitem[{{Wang} {et~al.}(2010){Wang}, {Carilli}, {Neri}, {Riechers}, {Wagg},
  {Walter}, {Bertoldi}, {Menten}, {Omont}, {Cox}, \&
  {Fan}}]{2010ApJ...714..699W}
{Wang}, R., {et~al.} 2010, \apj, 714, 699

\bibitem[{{Wang} {et~al.}(2013){Wang}, {Wagg}, {Carilli}, {Walter}, {Lentati},
  {Fan}, {Riechers}, {Bertoldi}, {Narayanan}, {Strauss}, {Cox}, {Omont},
  {Menten}, {Knudsen}, {Neri}, \& {Jiang}}]{2013arXiv1302.4154W}
---. 2013, arXiv:1302.4154

\bibitem[{{Wyithe}(2004)}]{2004MNRAS.351.1266W}
{Wyithe}, J.~S.~B. 2004, \mnras, 351, 1266

\bibitem[{{Zafar} {et~al.}(2010){Zafar}, {Watson}, {Malesani}, {Vreeswijk},
  {Fynbo}, {Hjorth}, {Levan}, \& {Micha{\l}owski}}]{2010A&A...515A..94Z}
{Zafar}, T., {Watson}, D.~J., {Malesani}, D., {Vreeswijk}, P.~M., {Fynbo},
  J.~P.~U., {Hjorth}, J., {Levan}, A.~J., \& {Micha{\l}owski}, M.~J. 2010,
  \aap, 515, A94

\end{thebibliography}

\clearpage



\clearpage

\clearpage

\end{document}